\documentclass[12pt, preprint]{aastex}

\shorttitle{}

\shortauthors{Jeltema et al.}

\begin{document}

\title{ The Evolution of Galaxies in X-ray Luminous Groups }

\author{Tesla E. Jeltema, John S. Mulchaey}

\affil{The Observatories of the Carnegie Institution of Washington}
\affil{813 Santa Barbara St., Pasadena, CA 91101}

\email{tesla@ociw.edu}

\author{Lori M. Lubin and Christopher D. Fassnacht}

\affil{Department of Physics}
\affil{University of California at Davis}
\affil{One Shields Ave., Davis, CA 95616}

\begin{abstract}

We investigate the galaxy populations in seven X-ray selected, intermediate-redshift groups ($0.2<z<0.6$).  Overall, the galaxy populations in these systems are similar to those in clusters at the same redshift; they have large fractions of early-type galaxies ($f_e \sim 70$\%) and small fractions of galaxies with significant star formation ($f_{[OII]} \sim 30$\%).  We do not observe a strong evolution in the galaxy populations from those seen in X-ray luminous groups at low-redshift.  Both $f_e$ and $f_{[OII]}$ are correlated with radius but do not reach the field value out to $\sim r_{500}$.  However, we find significant variation in the galaxy populations between groups with some groups having field-like populations.  Comparisons between the morphological and spectral properties of group galaxies reveal both gas-poor mergers and a population of passive spirals. Unlike low-redshift, X-ray emitting groups, in some of these groups the brightest galaxy does not lie at the center of the X-ray emission, and in several of the groups that do have a central BGG, the BGG has multiple components.  These groups appear to represent a range of evolutionary stages in the formation of the BGG.  Some groups have relatively large central galaxy densities, and one group contains a string of seven bright galaxies within a radius of 200 kpc that have a lower velocity dispersion than the rest of the system.  None of the central galaxies, including those with multiple components, have significant [OII] emission.  These observations support a scenario in which BGGs are formed relatively late through gas-poor mergers.

\end{abstract}


\section{ INTRODUCTION }

The morphologies and star-formation histories of galaxies depend on both environment and cosmic time.  Considerable
effort has gone in to tracing the correlation of morphology and star-formation rate with both local and global
environment in massive clusters, as well as the evolution of these properties with redshift (e.g. Dressler 1980; 
Butcher \& Oemler 1984; Dressler et al. 1997; Balogh et al. 1998, 1999; Poggianti et al. 1999; van Dokkum et al. 2000; 
Postman, Lubin, \& Oke 1998, 2001; Treu et al. 2003; Dressler et al. 2004; Postman et al. 2005; Tran et al. 2005a; 
Poggianti et al. 2006 to name a few). 
For example, in a recent paper Poggianti et al. (2006) find for one of the largest uniformly selected high-redshift
cluster samples yet studied, the optically-selected EDisCS clusters, that high-redshift clusters have a higher fraction
of star-forming galaxies than low-redshift clusters.  Overall, star formation in the universe has decreased significantly 
with time (e.g. Lilly et al. 1996; Madau et al. 1998; Hopkins 2004; Schiminovich et al. 2005).  Galaxies in 
clusters are known to have suppressed star-formation versus galaxies in the field at all redshifts (e.g. Balogh et al. 
1997, 1999; Poggianti et al. 1999; Postman et al. 2001; Lewis et al. 2002; Gomez et al. 2003; Kauffmann et al. 2004), 
but relatively few 
galaxies reside in such dense environments even at low redshift.  However, as many as 50\%-70\% of all galaxies reside 
in less massive groups of galaxies (Turner \& Gott 1976; Geller \& Huchra 1983; Eke et al. 2005), meaning that the 
group 
environment can play an important role in the evolution of galaxies.  Tracing the properties of galaxies versus both 
the 
local environment of the surrounding few galaxies and the global group/cluster environment has important implications 
for 
the processes responsible for both the morphological transformation
of galaxies and the reduction in star formation.  In the lower density environment of groups processes like 
ram-pressure 
stripping (e.g. Gunn \& Gott 1972; although see also Rasmussen, Ponman, \& Mulchaey 2006) are expected to be relatively 
ineffective, while other processes like galaxy-galaxy 
interactions and mergers are more effective (Barnes 1985; Aarseth \& Fall 1980; Merritt 1984, 1985; Zabludoff \& 
Mulchaey
1998; Miles et al. 2004; Taylor \& Babul 2005; Temporin \& Fritze-von Alvensleben 2006).

Studies of groups at low redshift have revealed them to be a heterogeneous population with their galaxy populations 
varying from cluster-like to field-like (Zabludoff \& Mulchaey 1998).  Groups showing extended X-ray emission tend
to have a significant fraction of early-type galaxies and a dominant early-type galaxy at the group center
(Mulchaey \& Zabludoff 1998; Mulchaey et al. 2003; Osmond \& Ponman 2004).  The existence of diffuse X-ray emission 
indicates a collapsed system with a relatively deep potential well, and these groups form the intersection between
less massive groups of a few galaxies, which may still be collapsing, and rich clusters.  These systems are also 
important to the understanding of processes like non-gravitational heating and cooling that affect the 
intracluster medium (e.g. Ponman, Sanderson, \& Finoguenov 2003 and references therein).

Given their heterogeneous nature and the large range of masses/densities represented among groups, sample selection
becomes a critical issue.  However, their galaxy densities make these systems difficult to recognize at even moderate
redshifts, and studies of representative samples of groups are just beginning.  Pushing toward lower densities,
Balogh et al. (2002a,b) studied galaxy morphologies and star formation in low-luminosity clusters at $z\sim0.25$.  They
find that the galaxy populations in these systems are similar to those in massive clusters at the same redshift.
On the optical side, Wilman et al. (2005a,b) use groups selected from the Canadian Network for Observational 
Cosmology 2 (CNOC2) survey to study star formation in groups at $0.3 < z \leq 0.55$.  Comparing these to 
optically-selected groups at low redshift, they find that the fraction of star-forming galaxies increases with
redshift in both groups and the field, but that this fraction is always lower in groups at a given epoch.  
Finally, moderate-redshift groups are being selected through their association with strong gravitational lenses 
(e.g., Fassnacht \& Lubin 2002; Momcheva et al. 2006; Auger et al. 2006), and the number of well-studied
groups found by this method is starting to approach those found by other techniques.

X-ray selection offers a natural method of identifying groups at intermediate redshifts as it confirms the presence
of a collapsed system.  This paper is part of a continuing multiwavelength project to study in detail the properties 
of X-ray selected groups at intermediate redshifts ($0.2<z<0.6$).  Our study represents an extension to both 
higher redshifts and lower masses of the study of low-luminosity clusters done by Balogh et al. (2002a,b), and it 
represents
a study of the X-ray emitting population of groups that can be compared to the optically-selected groups in Wilman 
et al. (2005a,b).  Mulchaey et al. (2006; Paper I) presented the spectroscopic confirmation of 
a sample of nine groups based on observations made with the Palomar 200 inch (5.08 m), Las Campanas 100 inch (2.54m), and Keck 10-m telescopes.  
Based on the
initial group membership these groups were found to have a significant fraction of early-type galaxies similar to X-ray
emitting groups at low redshift.  However, unlike low-redshift groups a central, dominant early-type galaxy was not 
found in $\sim40$\% of these systems.  In addition, three of the four central galaxies were found to have multiple 
components.  These observations suggest that X-ray emitting groups at intermediate redshift may be in an earlier phase of 
evolution where the central galaxy has not yet undergone its last major merger.  Jeltema et al. (2006; Paper II) presented
the X-ray properties of these groups based on observations taken with \textit{XMM-Newton}.  These observations, on the other hand,
confirm that these systems contain luminous, diffuse gas with luminosities and temperatures consistent with the
correlations seen among groups and clusters at low redshift.  The X-ray properties suggest that the group potential
is already largely in place.

In this paper, we present the results of deeper spectroscopic follow-up of seven groups with the Keck and Gemini
telescopes.  In combination with our \textit{Hubble Space Telescope (HST)} imaging, we study the dynamics, morphologies, and star-formation 
properties of X-ray selected, intermediate-redshift groups.  In \S2, we present the observations and data reduction
as well as the spectroscopic catalog.  \S3 discusses the group velocity dispersions and their relationship to the
X-ray properties.  The morphological composition and star-formation in these groups compared to low-redshift 
groups and clusters are presented in \S4 and \S5, respectively.  Finally, in \S6 we revisit the properties of
the brightest group galaxies (BGGs). Throughout the paper, we assume a cosmology 
of $H_0=70h_{70}$ km s$^{-1}$ Mpc$^{-1}$, $\Omega_m=0.27$, and $\Lambda = 0.73$.

\section{ OBSERVATIONS AND DATA REDUCTION }

\subsection{ Group Sample }

As discussed in Papers I and II, our intermediate-redshift groups are X-ray selected; they 
are drawn from objects in the \textit{ROSAT} Deep Cluster Survey (RDCS;
Rosati et al. 1998) with luminosities in the group regime and redshifts greater than 0.2.
As shown in Paper I, the groups studied here are typically the more luminous RDCS groups, and
all but one have fluxes in the range where the RDCS is $\sim$90\% complete (Rosati et al. 1998).
In this paper, we discuss the seven northern groups in Paper I.  We have obtained 
deeper spectroscopy for these groups with the Keck and Gemini-North telescopes.  For six groups,
we also have X-ray data from \textit{XMM-Newton} (see Paper II).

\subsection{ Spectroscopy }

We observed five groups with GMOS (Hook et al. 2004) on Gemini-North during 2005 May 11-13, while two groups, 
RXJ0329+02 and RXJ0720+71, were observed with LRIS (Oke et al. 1995; McCarthy et al. 1998) on Keck I during 
2004 December 5-6.  Spectroscopic targets 
for the Gemini observations were selected from GMOS pre-imaging of each field (see \S2.3).
Only galaxies with R-band magnitudes brighter than 21.5 in the Gemini imaging were selected and 
priority was given to 
galaxies with magnitudes brighter than 20.5.  No color or morphology information was used in the selection.
We observed two to four masks per group with 12-20 slits per
mask.  GMOS was operated with the B600 G5303 grating with a central wavelength of 5500 \AA\ 
leading to a spectral coverage of $\sim 3500-7000$ \AA.  Spectra were binned to a pixel scale
of 1.8 \AA\ per pixel to improve the signal-to-noise ratio (S/N).  The chosen 1'' slit width gives a spectral resolution of
approximately 5.5 \AA.  
Two to three 30 minute exposures were taken per mask.

The Gemini spectra were reduced using the Gemini IRAF package.  The reduction included bias subtraction,
flat-field correction, cosmic-ray rejection, sky subtraction, wavelength calibration, and extraction
of one-dimensional (1D) spectra.  The 1D spectra of multiple science exposures were combined after the 
rejection of outlying pixels using the averaged sigma clipping algorithm.

With LRIS, we observed two masks each for RXJ0329+02 and RXJ0720+71 with three to four half hour exposures.  Spectroscopic
targets were selected from the RDCS imaging (Rosati et al. 1998) which was in I-band for RXJ0329+02 and in R-band for 
RXJ0720+71.  Galaxies were selected within the magnitude ranges 18-22 for RXJ0329+02 (I-band) and 17-23 for 
RXJ0720+71 (R-band), and they were given priority based on both magnitude and distance from the BGG (i.e. 
we preferentially targeted closer and brighter galaxies).  Again no color or morphology information was used 
in the selection.  On the
blue side the 600/4000 grism was used for both groups.  On the red side, we chose the 600/5000 grating 
for RXJ0329+02 and the 600/7500 grating for RXJ0720+71.  For RXJ0329+02, a 4600 \AA\ dichroic was chosen 
such that all interesting features fall on the red side for this group, and only these data were used in the 
analysis.  The red side data has a pixel scale of 1.28 \AA\ per pixel, and the spectral coverage for 
RXJ0329+02 is roughly 4800-7700 \AA.  For RXJ0720+71, a 5600 \AA\ dichroic was chosen.  For this group
[OII] emission and the important absorption features (through the G-band) fall on the blue side.  The 
initial pixel scale of the blue side was 0.63 \AA\ per pixel, so these data were binned by three to improve
the S/N and to better match the pixel scale of the Gemini data.  Here the redshift fitting was 
restricted to the 4000-5600 \AA\ range.  The LRIS data were reduced with custom scripts that serve as front ends
to the standard IRAF tasks (M. Auger, private communication 2006).

We determine galaxy redshifts using the cross-correlation routine XCSAO (Kurtz \& Mink 1998) 
after the removal of very bad
sky and cosmic ray features.  Spectra are correlated with
the galaxy and quasar templates from the Sloan Digital Sky Survey (SDSS) Data Release 4.  All redshifts 
were confirmed by eye by TEJ.  Only fits with $R \geq 3$ were considered.  Redshift errors are estimated
by XCSAO based on the width of the correlation peak (Tonry \& Davis 1979; Kurtz \& Mink 1998).  Kurtz \& 
Mink (1998) compare the XCSAO error estimates to the differences in redshift found from two observations
of the same object and find that the distribution of XCSAO errors is similar within about 20\%.  As a 
second indication of the redshift errors, we measured redshifts separately for the three exposures of mask
four for RXJ1648+60.  This mask was observed over two different nights during the Gemini run.  We find 
that the offsets in redshift determined from the individual exposures
compared to the combined spectrum are similar to the errors estimated by XCSAO so we adopt the errors from XCSAO.

The Gemini spectra were taken under excellent conditions (seeing of 0.35''
to 0.65''), and 94\% of the targeted galaxies yielded redshifts.  The Keck data were taken under 
significantly poorer conditions with seeing between 1'' and 2'', and 52\% of the targeted galaxies yielded redshifts.  
These data were judged to be sufficient to determine galaxy redshifts, but due to the generally poorer
quality of the Keck data compared to the Gemini data we did not determine equivalent widths for these spectra. 

For the Gemini groups, our spectroscopy is 45\%-90\% complete to $M_V=-20.5 + 5$log$_{10} h_{70}$ within a radius of 
700 $h_{70}^{-1}$ kpc 
(the largest radius probed by the spectroscopy in all groups).  This magnitude limit corresponds to 
approximately $M_V^*+1$ (Dressler et al. 1999).  Some groups have good completeness to significantly lower
 magnitudes due to the variation in redshift among the sample and the number of masks observed, but a magnitude limit 
of $M_V=-20.5 + 5$log$_{10} h_{70}$ yields reasonable completeness for the entire sample.  Because we obtain a redshift for 
nearly every galaxy targeted, we do not expect any bias toward emission-line galaxies.  The Keck spectroscopy 
for RXJ0720+71 is also fairly complete (75\%) within the same radius and magnitude limits, but the 
completeness for RXJ0329+02 is somewhat lower at 32\%.  Here there may be some concern about a bias toward 
emission-line redshifts, but in fact the Keck spectra reveal very few galaxies with visible emission lines 
(4 out of 28 group members), and all but one of these spectra have visible absorption lines.  
In addition, in the discussion of galaxy morphology below all of the group galaxies in RXJ0329+02 and RXJ0720+71 
covered by 
the HST imaging turn out to be early-type galaxies, and these groups are not considered in our discussion of [OII] emission.

\subsection{ Imaging }

As discussed in Paper I, R-band imaging for these groups covering the full field of the spectroscopy was 
obtained using the Palomar 200-inch and Las Campanas 100-inch telescopes.  This imaging was used to both 
classify objects as galaxies or stars and to measure magnitudes.  To minimize the K-corrections, we convert
the observed R-band magnitudes to rest frame V-band absolute magnitudes; this conversion is done using the
spectral energy distributions of Coleman, Wu, \& Weedman (1980). 
Higher resolution images of the centers of the groups were taken with WFPC2 on the 
Hubble Space Telescope using the F702W filter with exposure times ranging from 4400 to 10,400 secs.  
From the HST imaging, we determine galaxy morphology for all galaxies with measured 
redshifts that fall within the WFPC2 field of view.  Galaxies were visually classified by T.E.J. and L.M.L. and 
independently by J.S.M.  These independent classifications only disagreed for 12\% of the galaxies, and for 
only one of the 112 galaxies classified was the disagreement greater than one morphological class.

For the five groups observed with Gemini, additional imaging was taken as pre-imaging for the spectroscopic
observations.  Images were taken using the r\_G0303 filter for total exposure times of 8-9 minutes and with 
seeing ranging from 0.6'' to 1.1''.  These data were judged to be sufficient to classify galaxies as ``early''
versus ``late'' types, allowing a rough morphological classification for almost all galaxies with measured
redshifts.  Here the disagreement in classification between independent investigators was 15\%.  For those
galaxies with both HST and Gemini imaging, the classifications differ for 24\% of the galaxies.  The HST data
reveals that in most of these cases a late-type galaxy was mistakenly classified as an early-type galaxy
from the ground-based imaging due to the poorer resolution and loss of low surface brightness disk features in 
the ground-based data.  Early-type fractions based on the Gemini data should therefore be considered
upper limits.  Where available we use the HST morphologies.

\subsection{Spectroscopic Catalog}

Table 1 lists the complete spectroscopic catalog.  The columns are:

\noindent(1)  R.A. (J2000)\\
(2)  Dec. (J2000)\\
(3)  Redshift\\
(4)  Redshift error\\
(5)  Telescope used for redshift determination: Gemini (G), Keck (K), or Palomar (P)\\
(6)  R-band magnitude\\
(7)  Magnitude error\\
(8)  HST morphology\\
(9)  Early or Late type classification (Taken from the HST imaging where available and the Gemini imaging otherwise.)\\
(10)  Notes on morphology.

\section{ DYNAMICAL PROPERTIES }

\subsection{ Group Membership and Velocity Dispersion }

Figure 1 shows the distribution of measured redshifts for each of the group fields.  Where available we use
the redshifts from the Gemini or Keck data, but we also include any additional galaxies with known 
redshifts determined from our Palomar data in Paper I.  In most cases, the group is obvious as the most 
significant peak, but in a few fields there is more than one significant galaxy system.  In particular, 
the RXJ1205+44 field contains two foreground groups at $z = 0.338$ and $z = 0.373$; however, the XMM-Newton 
observation (Paper II) reveals that the X-ray emission is clearly peaked on a bright galaxy in the 
system at $z=0.593$.  Several other galaxies in this highest-redshift system also fall within the X-ray
emission.  The RXJ1334+37 and RXJ1648+60 fields also show multiple possible galaxy associations; however, 
there is a clearly dominant peak in these galaxy distributions, and the X-ray emission is associated
with galaxies in these dominant galaxy systems.

As in Paper I, we determine group membership using the ROSTAT package (Beers, Flynn, and Gebhardt 1990).
Initially, all galaxies within $\pm0.01$ of the group mean redshift are considered, and we calculate the
biweight estimators of the mean group velocity and velocity dispersion.  We then exclude any galaxies with 
velocities offset from the mean by more than 3 times the velocity dispersion and repeat the 
calculations.  This procedure led to the removal of one galaxy from RXJ1256+25, but no galaxies were
excluded from the other groups.  The velocity distributions of the group members are also shown in Figure 1.
We identify between 10 and 33 members in these groups.
\clearpage
\begin{center}
\epsscale{1.0}
\plottwo{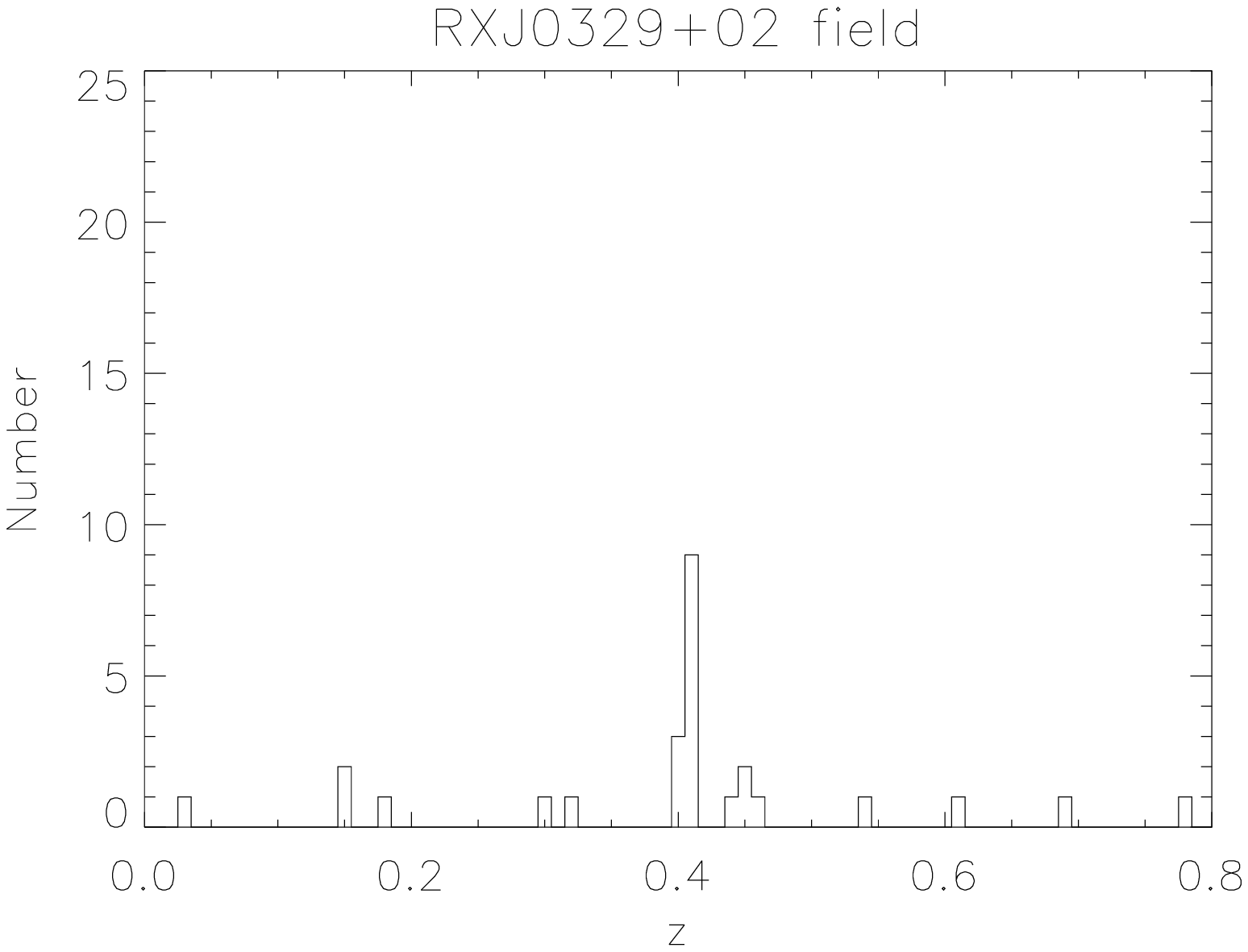}{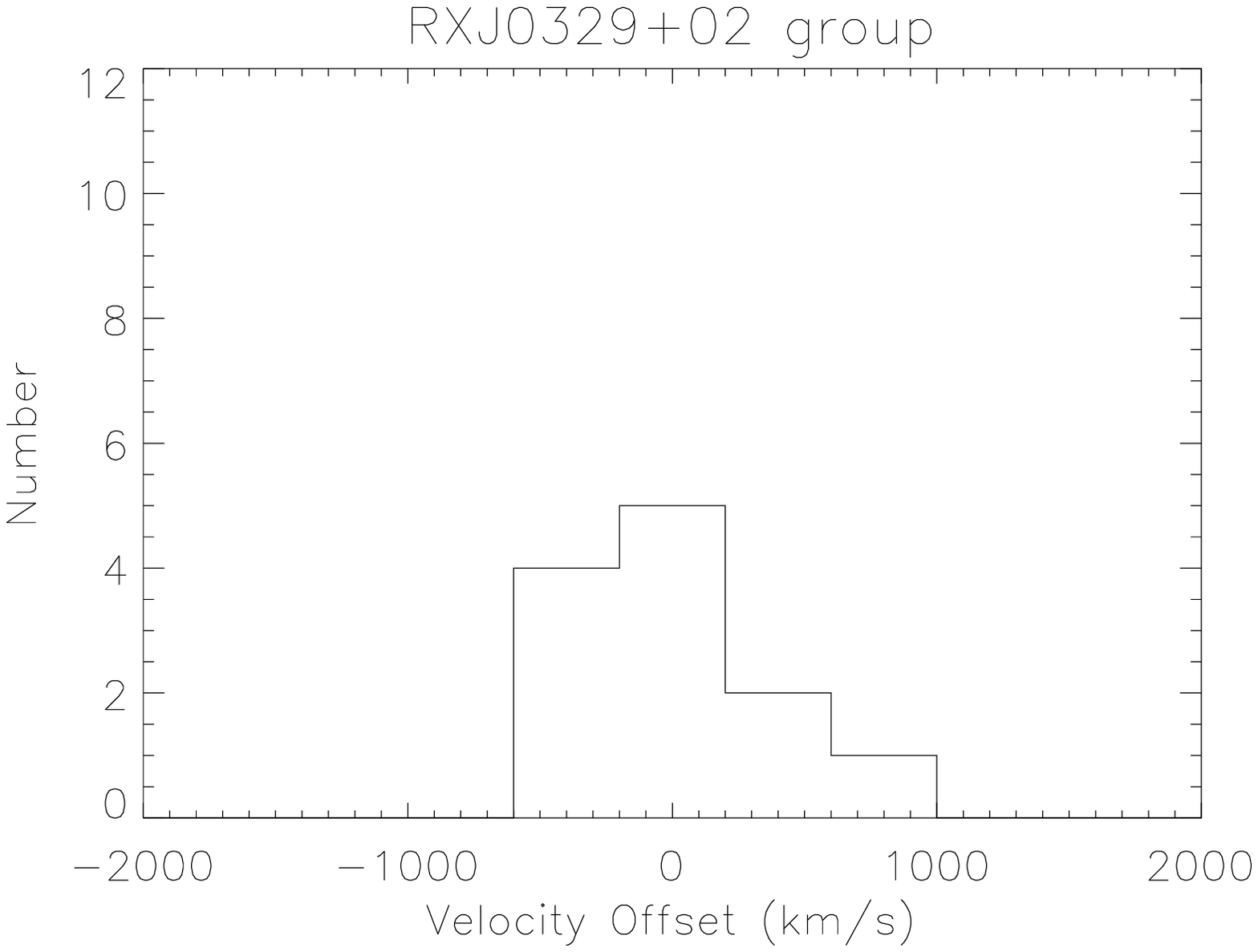}
\plottwo{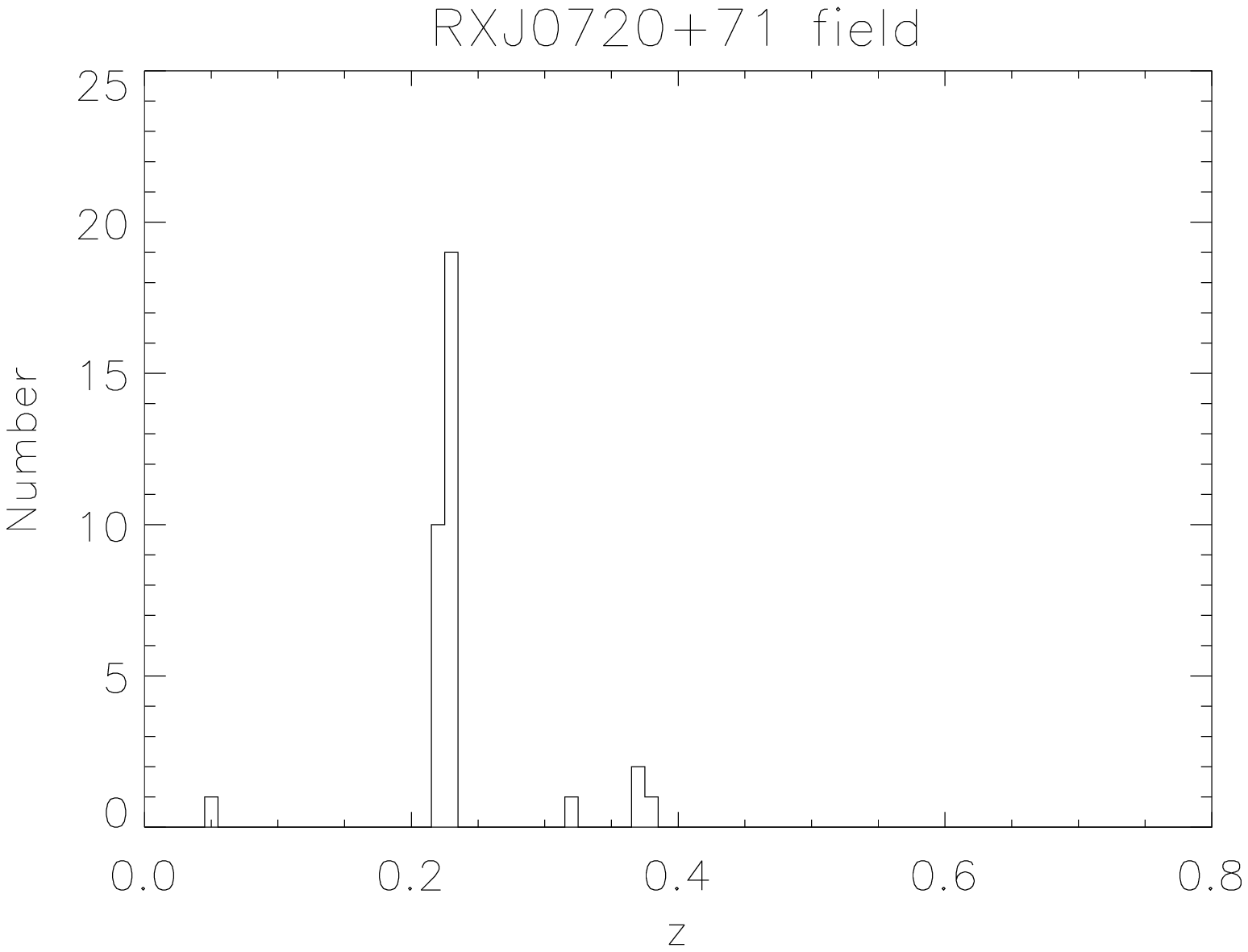}{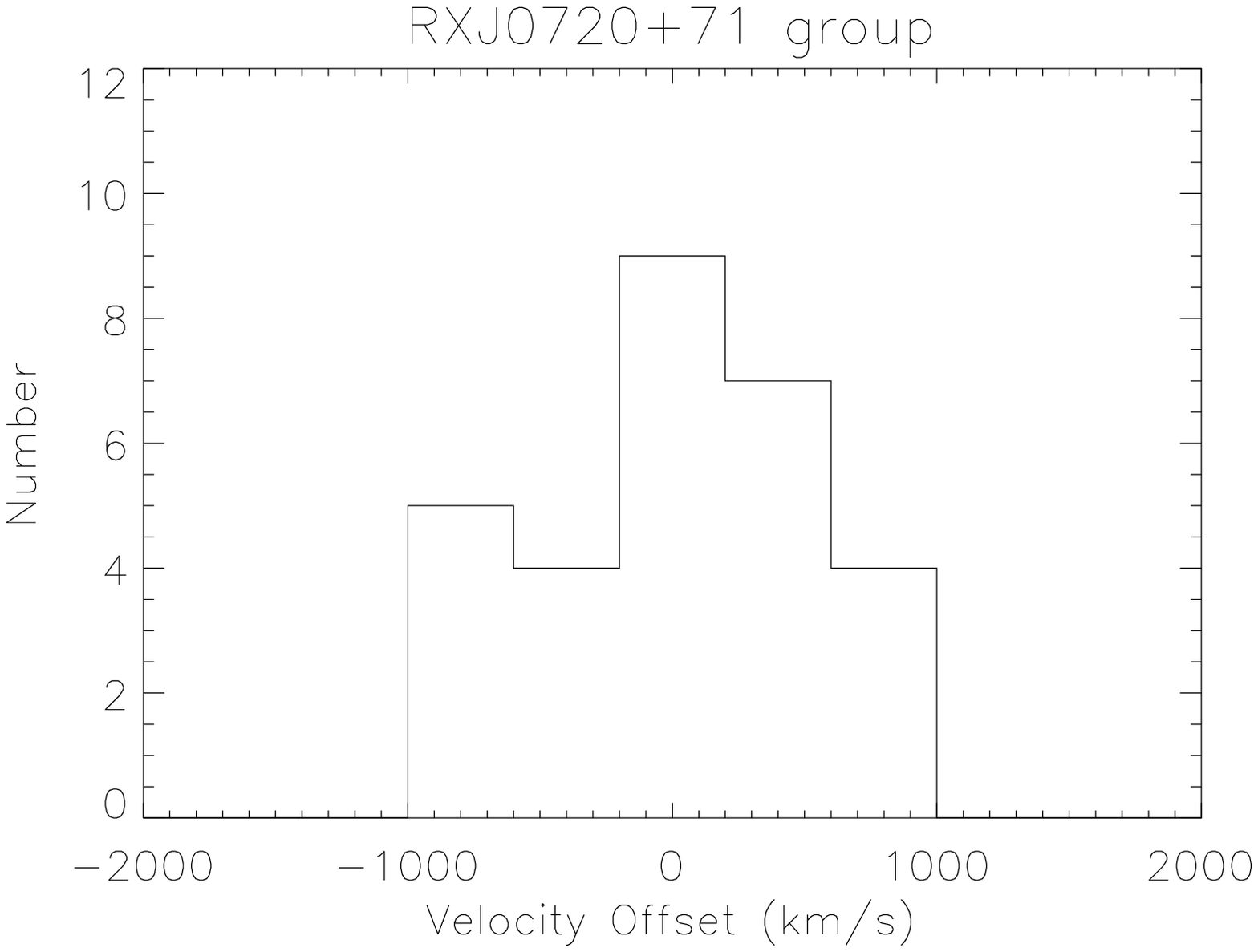}
\plottwo{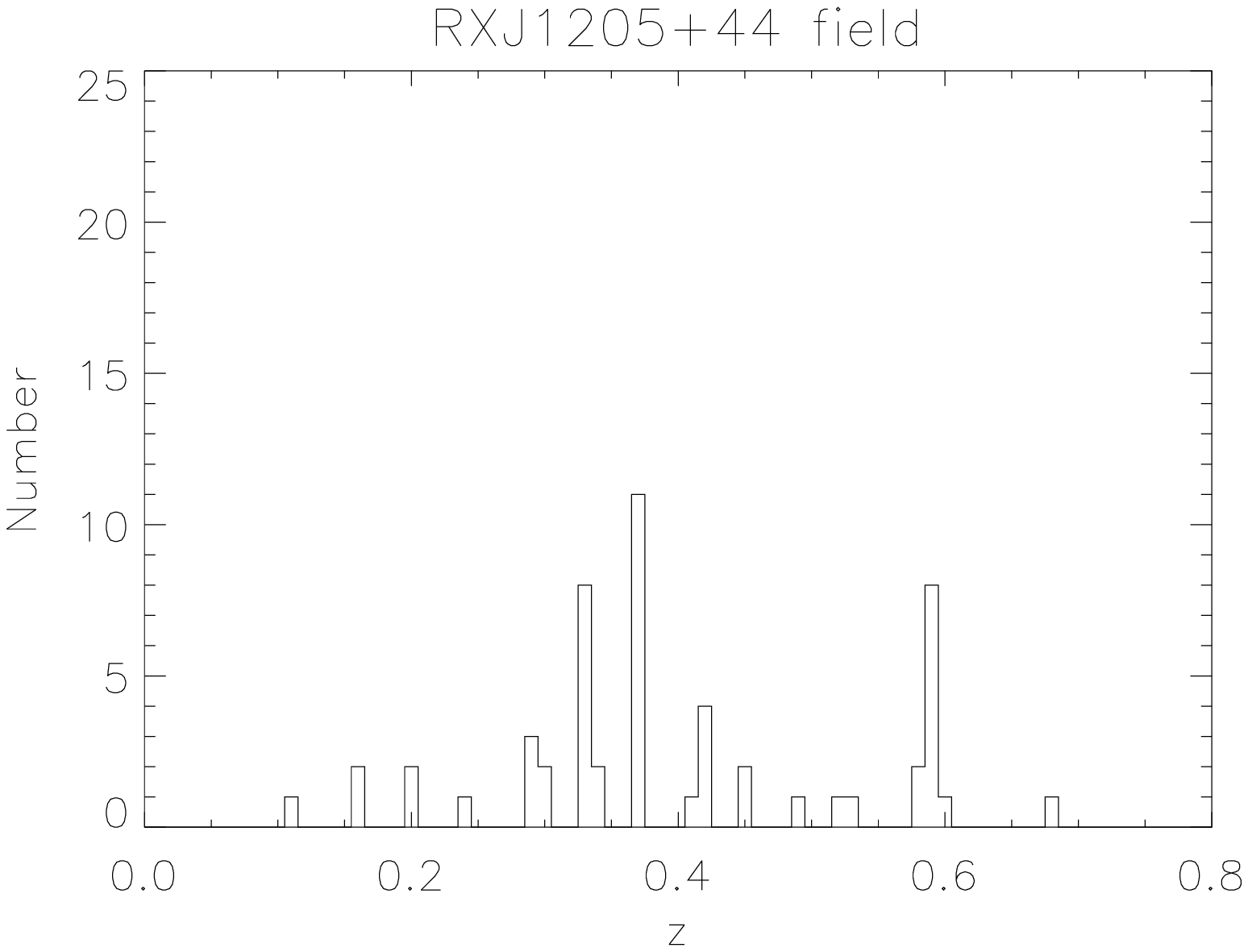}{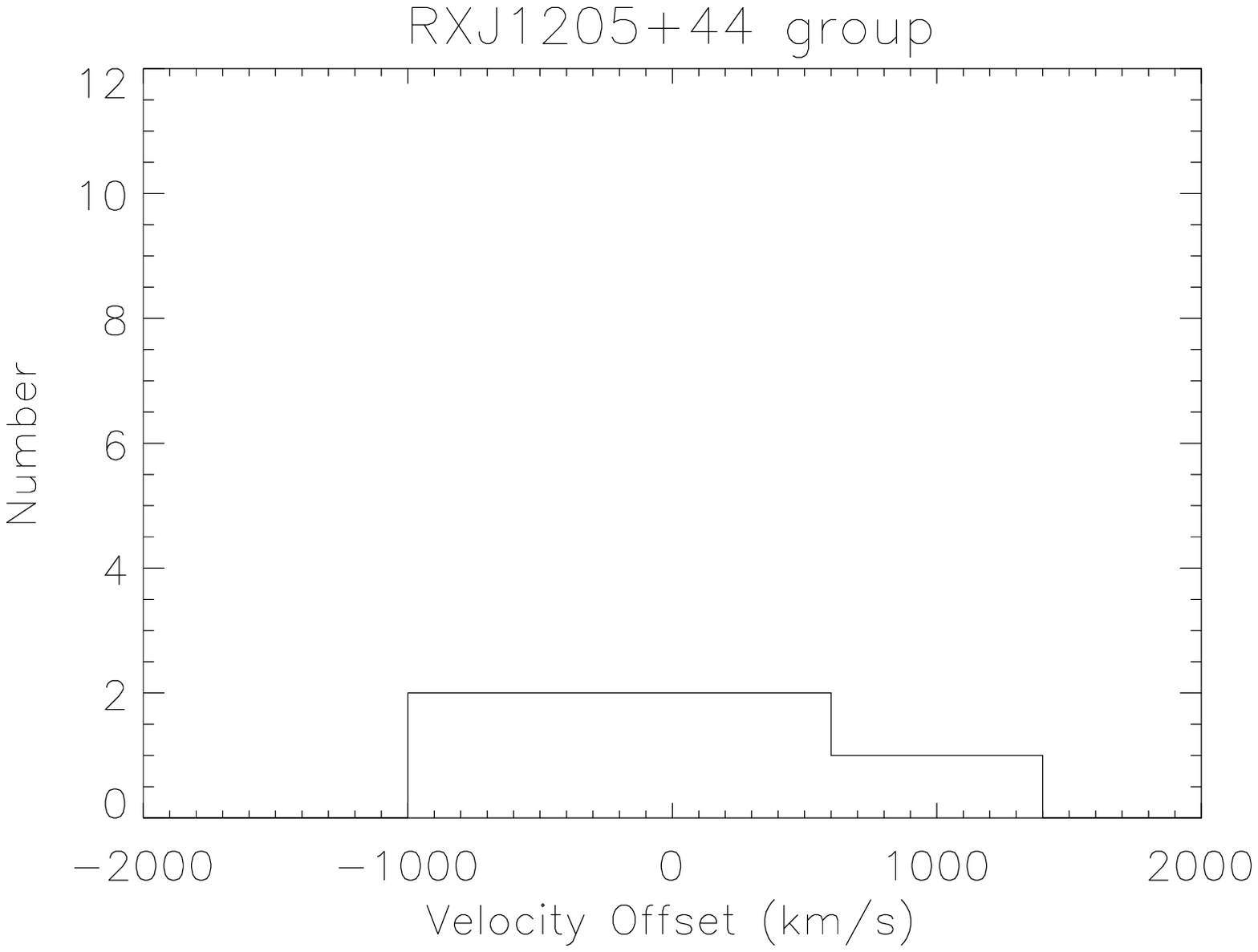}
\plottwo{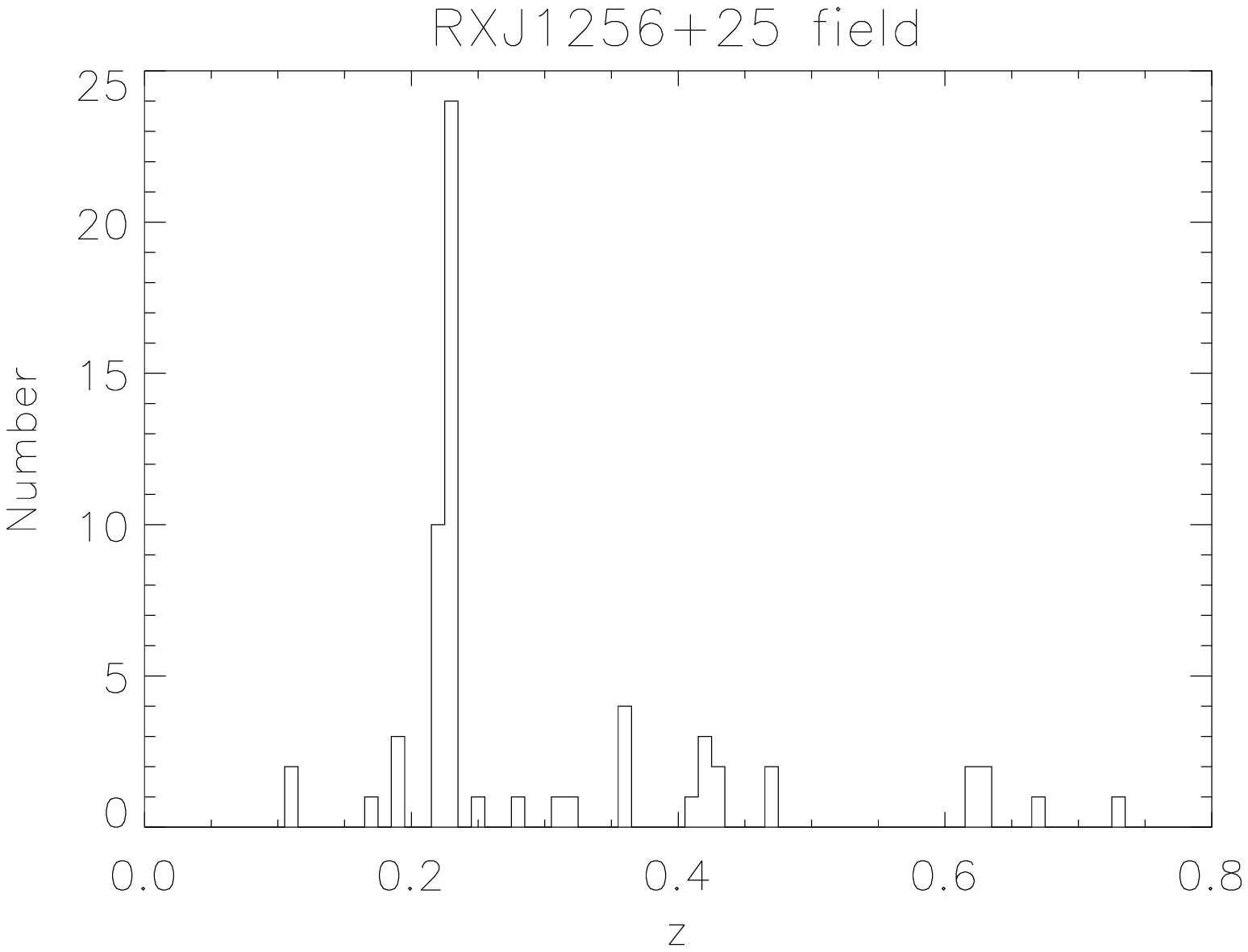}{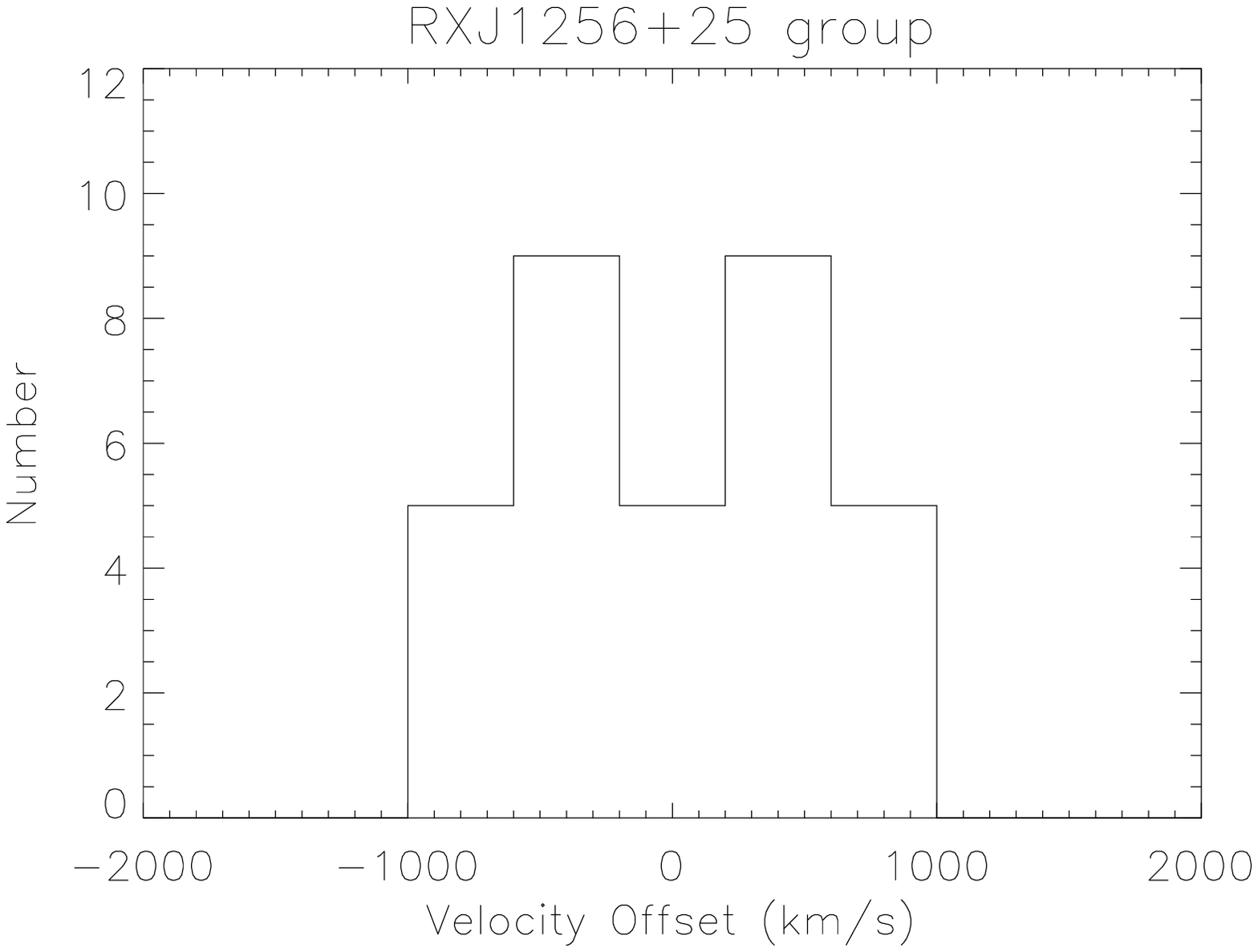}
\end{center}
\clearpage
\begin{figure}
\plottwo{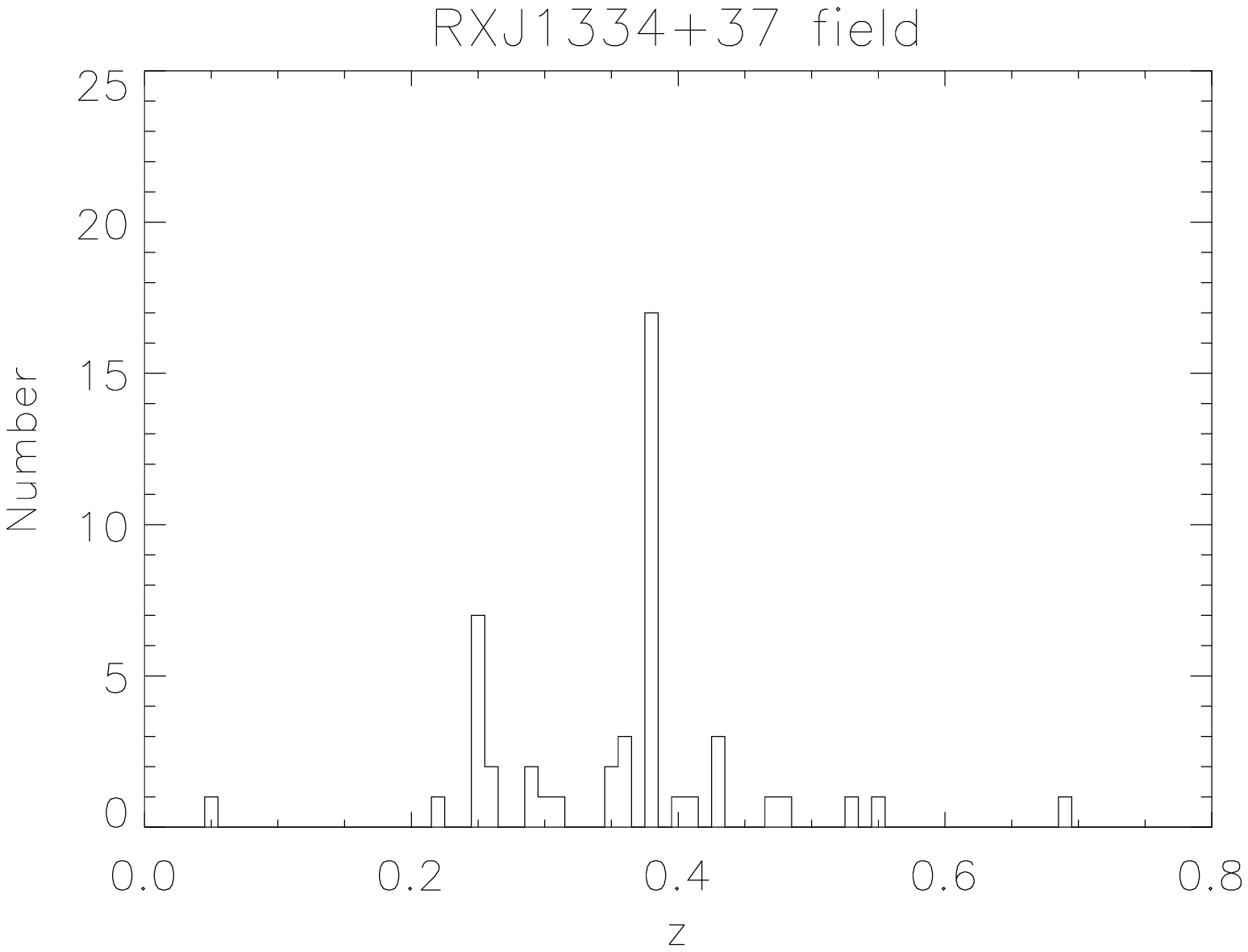}{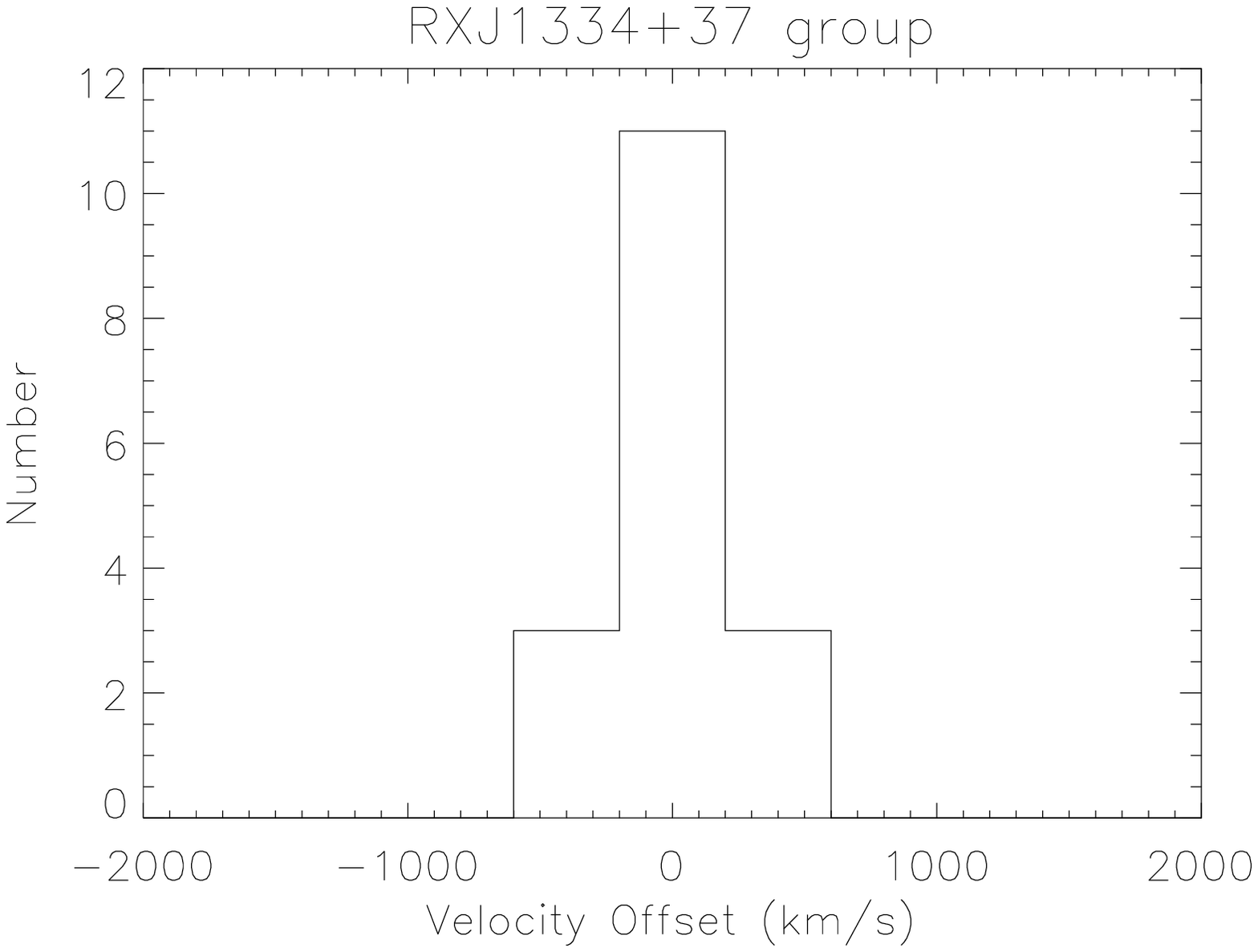}
\plottwo{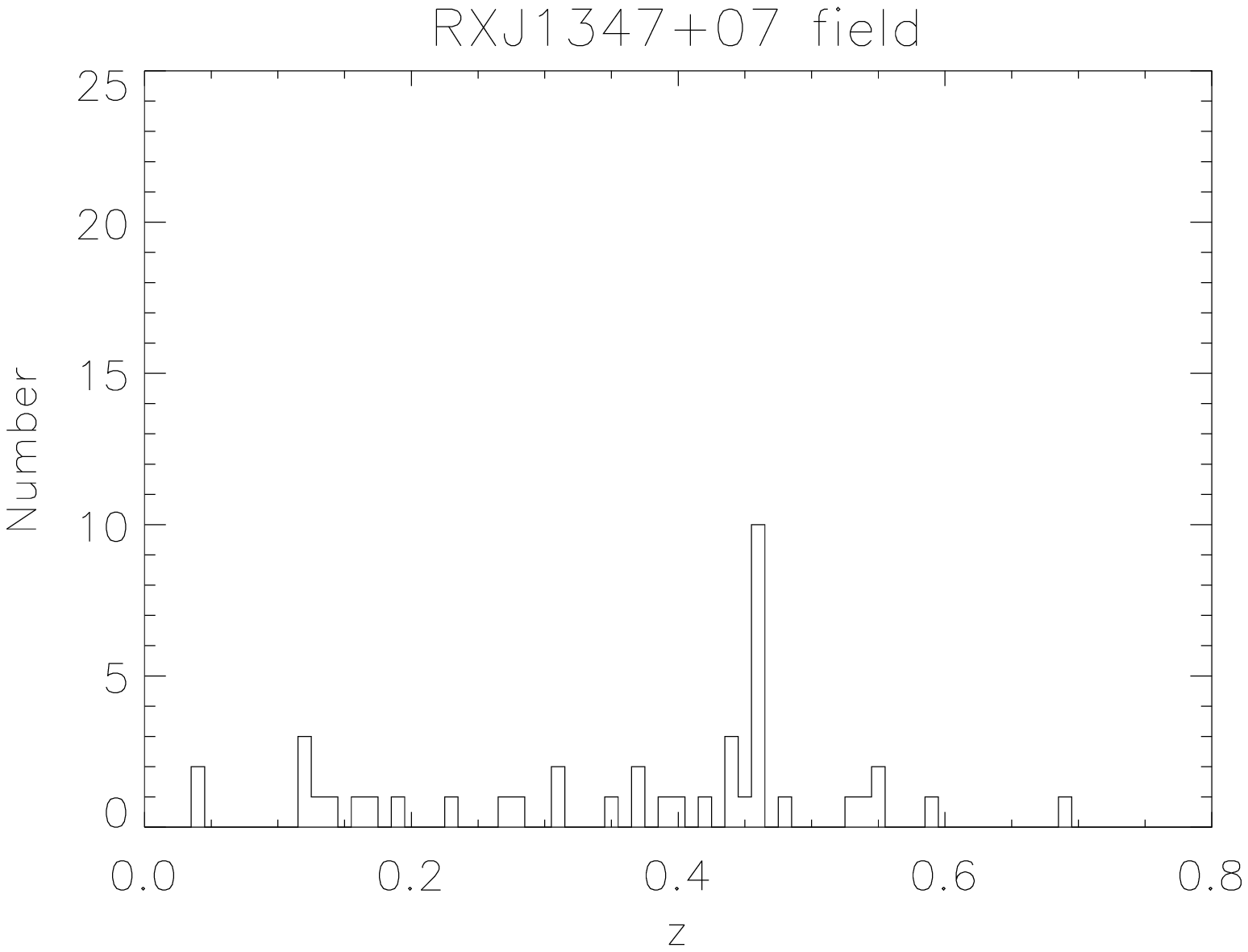}{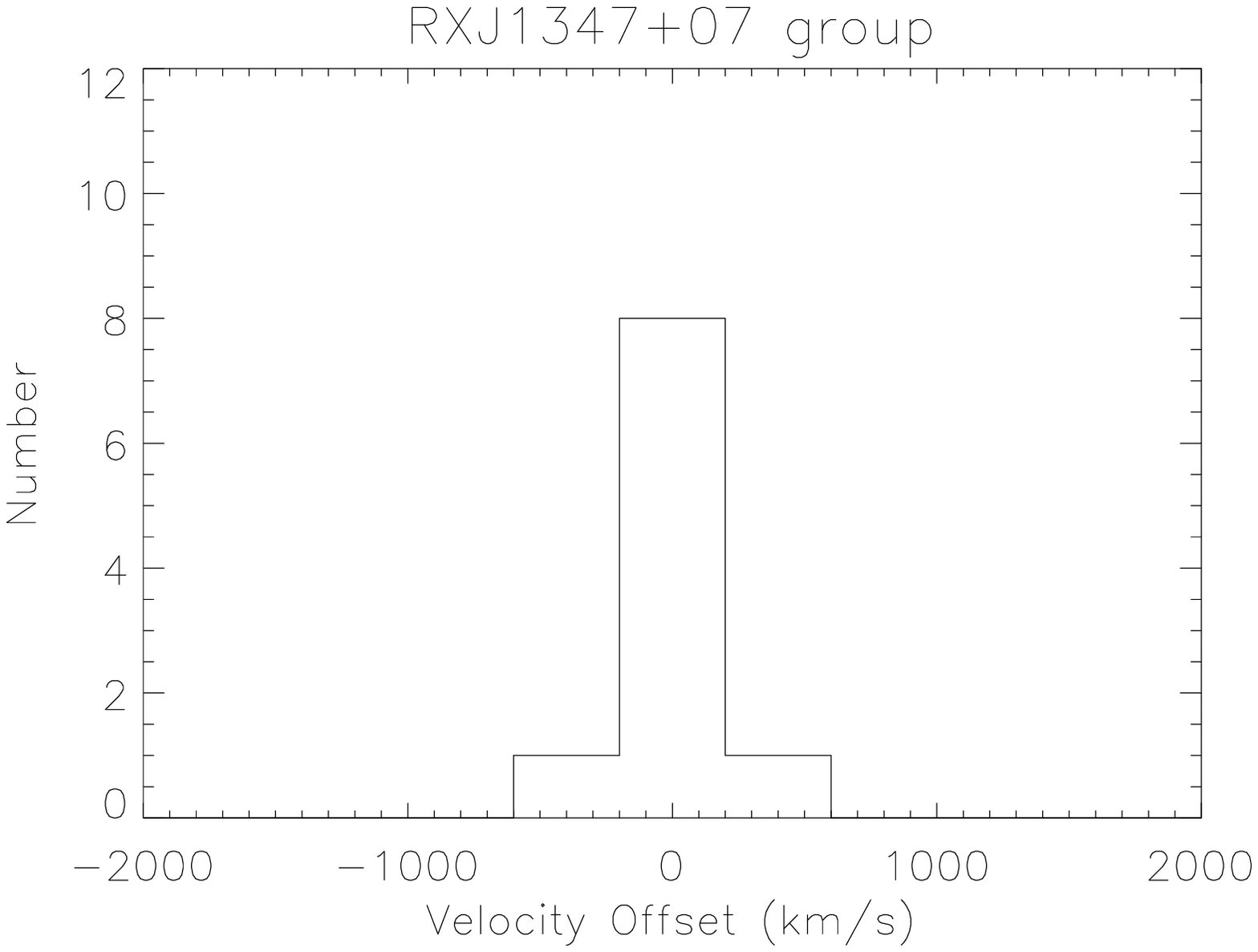}
\plottwo{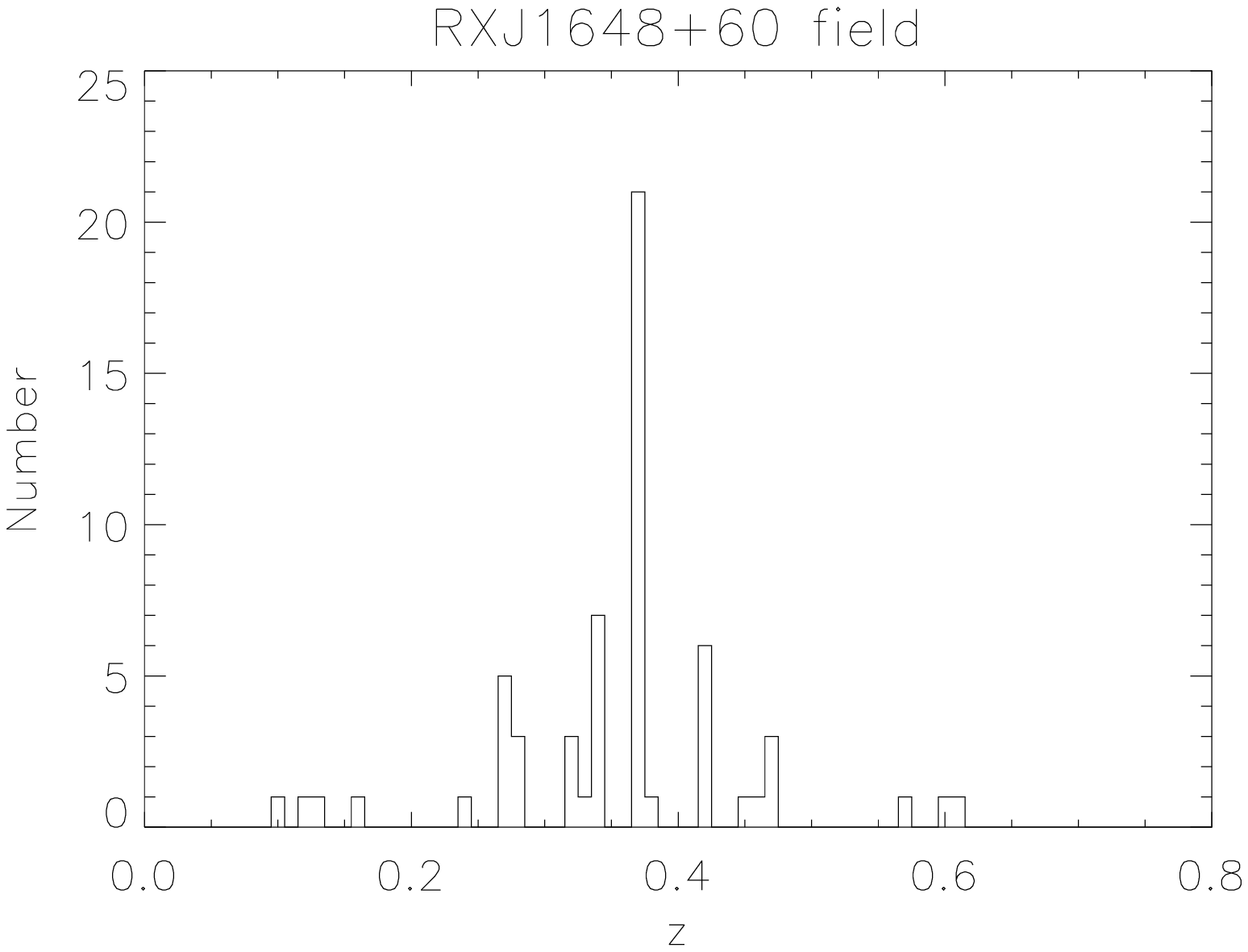}{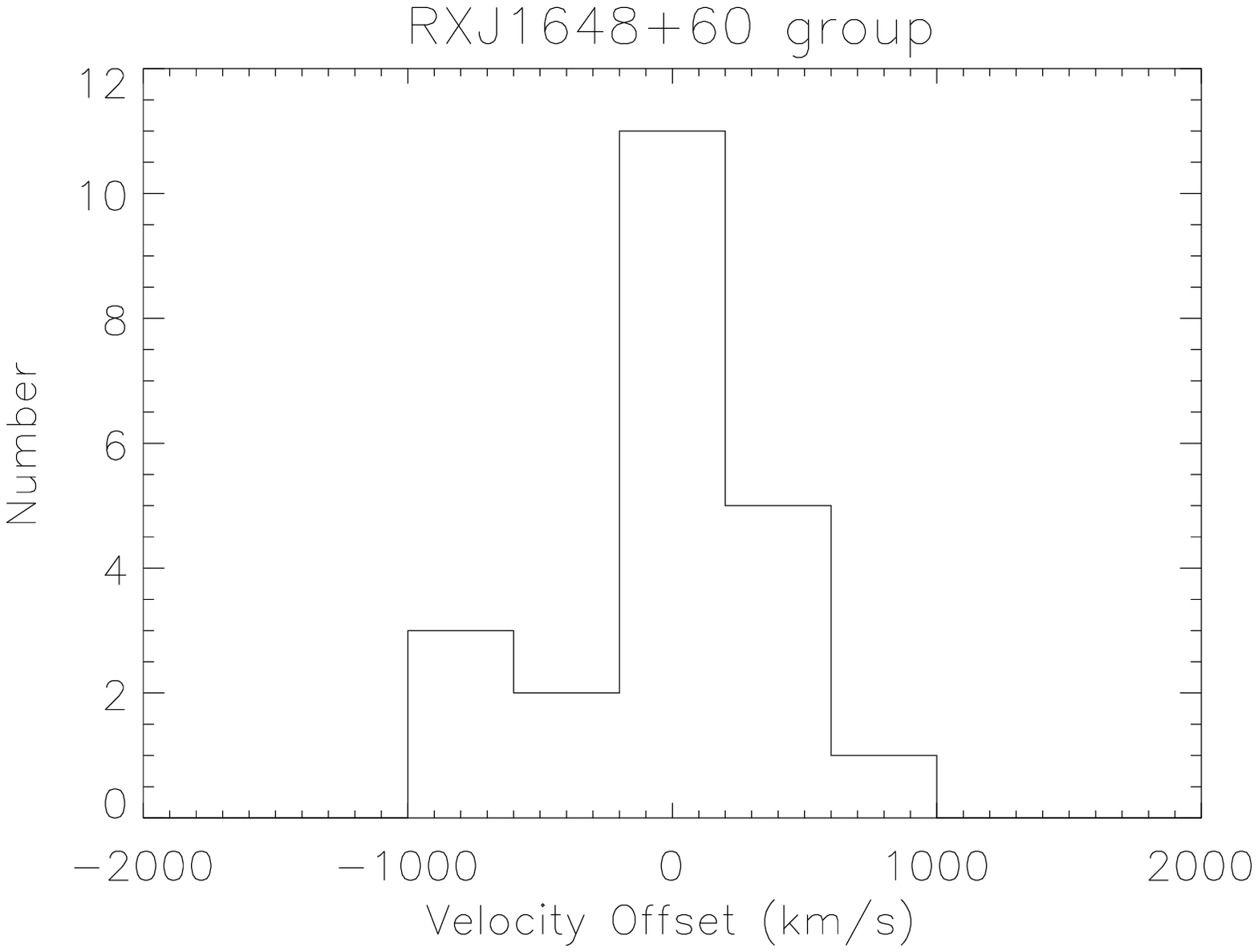}
\caption{ Right: Redshift distribution of the galaxies in each group field.  Left: Velocity distribution
of the group members in terms of the offset from the mean group velocity. }
\end{figure}
\clearpage
Table 2 lists the group mean redshifts and velocity dispersions 
determined from both all known members and the Gemini or Keck data alone.  For RXJ0329+02,
no Palomar data were available.  Errors in the velocity
dispersion are the $1 \sigma$ bootstrap errors on the biweight estimate of the velocity dispersion.
For RXJ1256+25, the velocity dispersion determined from the Gemini data is significantly lower than the
dispersion from all group members, but for all other groups these two measurements are consistent within
the errors.  The difference in velocity dispersion for RXJ1256+25 appears to be due to a tendency for
the galaxies with Palomar redshifts to be at lower redshifts.  For the 15 galaxies in this group for
which we have redshifts from both Gemini and Palomar, we find no significant tendency for the redshifts
from one telescope to be higher or lower than from the other.  However, the Palomar galaxies in this
group tend to fall in the northwest region of the group.  Considering only those group members with
Gemini redshifts, both a Wilcoxon rank-sum test (e.g., Walpole \& Myers 1993) and a Kolmogorov-Smirnov 
test (e.g., Press et al. 1992, \S 14.3) show a significant  tendency for galaxies in the northern
half of the group to have lower redshifts, possibly indicating substructure in this group.  The velocity distribution
for this group shown in Figure 1 does show two peaks; however, we do not find a significant deviation from 
normality in the velocity distribution.

Given the redshift range spanned by our group sample, our spectroscopy probes a varying physical radius
in these groups.  However, even in the lowest redshift groups we have galaxy redshifts out to radii of at least
 700 $h_{70}^{-1}$ kpc.  This radius corresponds to at least half of $r_{200}$ (determined from the X-ray data in
Paper II), and in all but the two lowest-redshifts groups we probe beyond $r_{200}$.  We find that 
recalculating the velocity dispersions within a fixed physical radius of 700 $h_{70}^{-1}$ kpc does not change the
velocity dispersions within the errors.
In the interest of improved statistics, in the rest of the paper we use the velocity dispersions from
all of the known members.

The velocity dispersions for both RXJ1334+37 and RXJ1648+60 increase significantly from those in
Paper 1.  For both of these groups we have now identified nearly 3 times as many members.
In the case of RXJ1648+60, the Palomar data in Paper I sampled only the brightest group members,
with seven of the eight brightest galaxies represented among the eight group members in Paper I.
This group does not contain a single bright central galaxy; instead it is centered on a string of
bright galaxies.  
In \S6, we show that these galaxies have a smaller velocity dispersion than the rest of the group.
In RXJ1334+37, there is no clear trend for the six members in Paper I to have different magnitudes or radii
than the galaxies in this paper, and here we may simply be seeing the effect of small sample size.
This group also shows some indications of substructure.  The Gemini spectroscopy reveals four group members 
approximately 2.5$'$ north of the group center that tend to have higher redshifts than the rest of the group;
the position of these galaxies corresponds to a faint, marginally significant excess in X-ray emission.
Removing these galaxies from the sample reduces the best-fit velocity dispersion to 185 km s$^{-1}$, which
is only a significant reduction at about 2 $\sigma$.

\subsection{ $L_X-\sigma$ and $\sigma-T$ }

Figure 2 shows the relationship between the velocity dispersions and the X-ray luminosities and 
temperatures of our groups compared to low-redshift groups and clusters.  These figures are the same as in 
Papers I and II 
updated with the new velocity dispersions.  The luminosities of the intermediate-redshift groups have been 
corrected by the evolutionary factor $E_z^{-1}L_X$, where
$E_z=H_z/H_0=[\Omega_m(1+z)^3+\Lambda]^{1/2}$.  With the increase in the velocity dispersions of RXJ1334+37 and 
RXJ1648+60 with respect to Paper I, our intermediate redshift groups generally agree quite well with the 
properties of the low-redshift systems, also showing similar scatter.  There is still some tendency for 
intermediate-redshift groups to have velocity dispersions that are low given their luminosities, but there is 
significant scatter in this relationship.
\clearpage
\begin{figure}
\epsscale{0.7}
\plotone{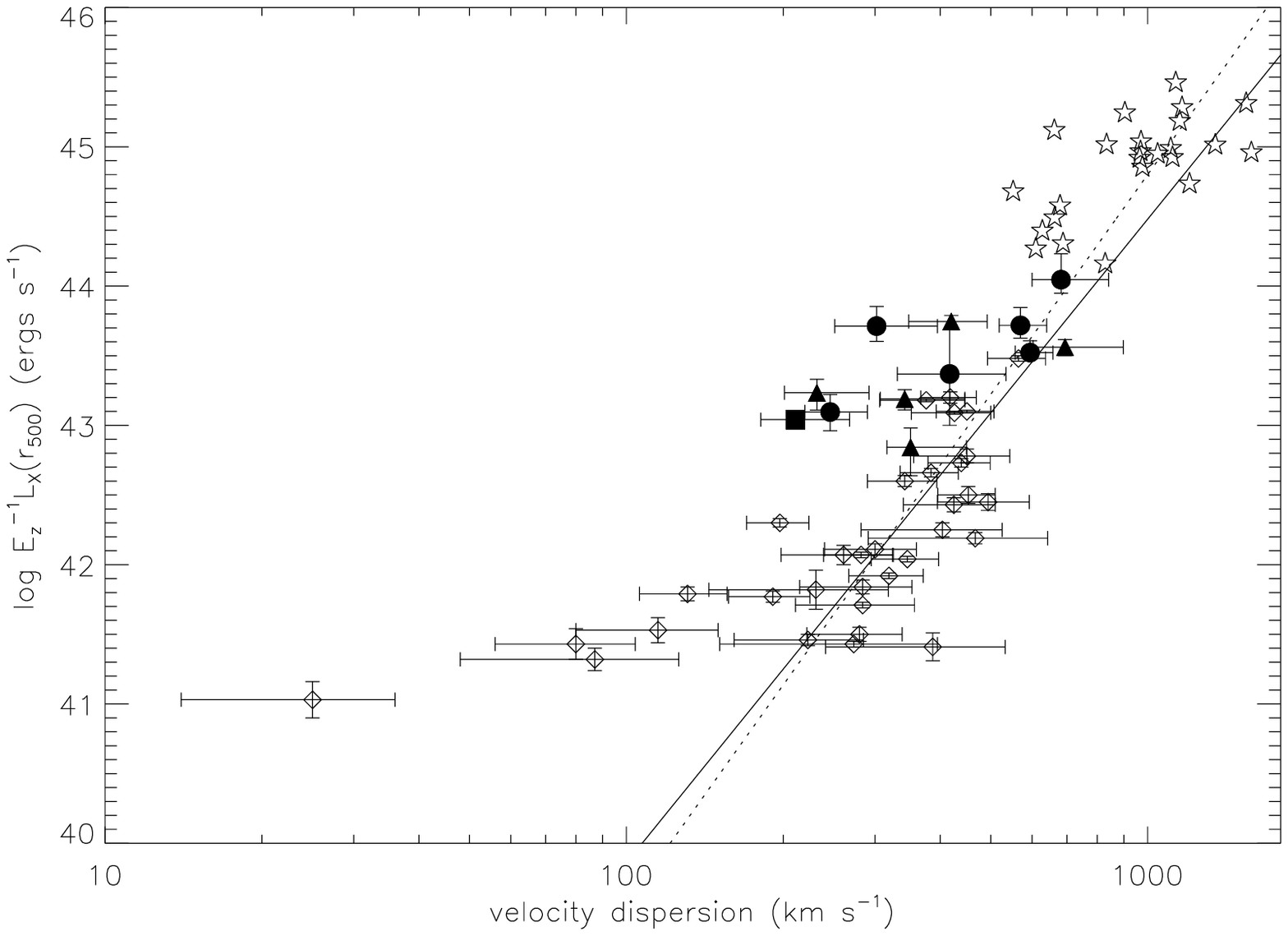}
\plotone{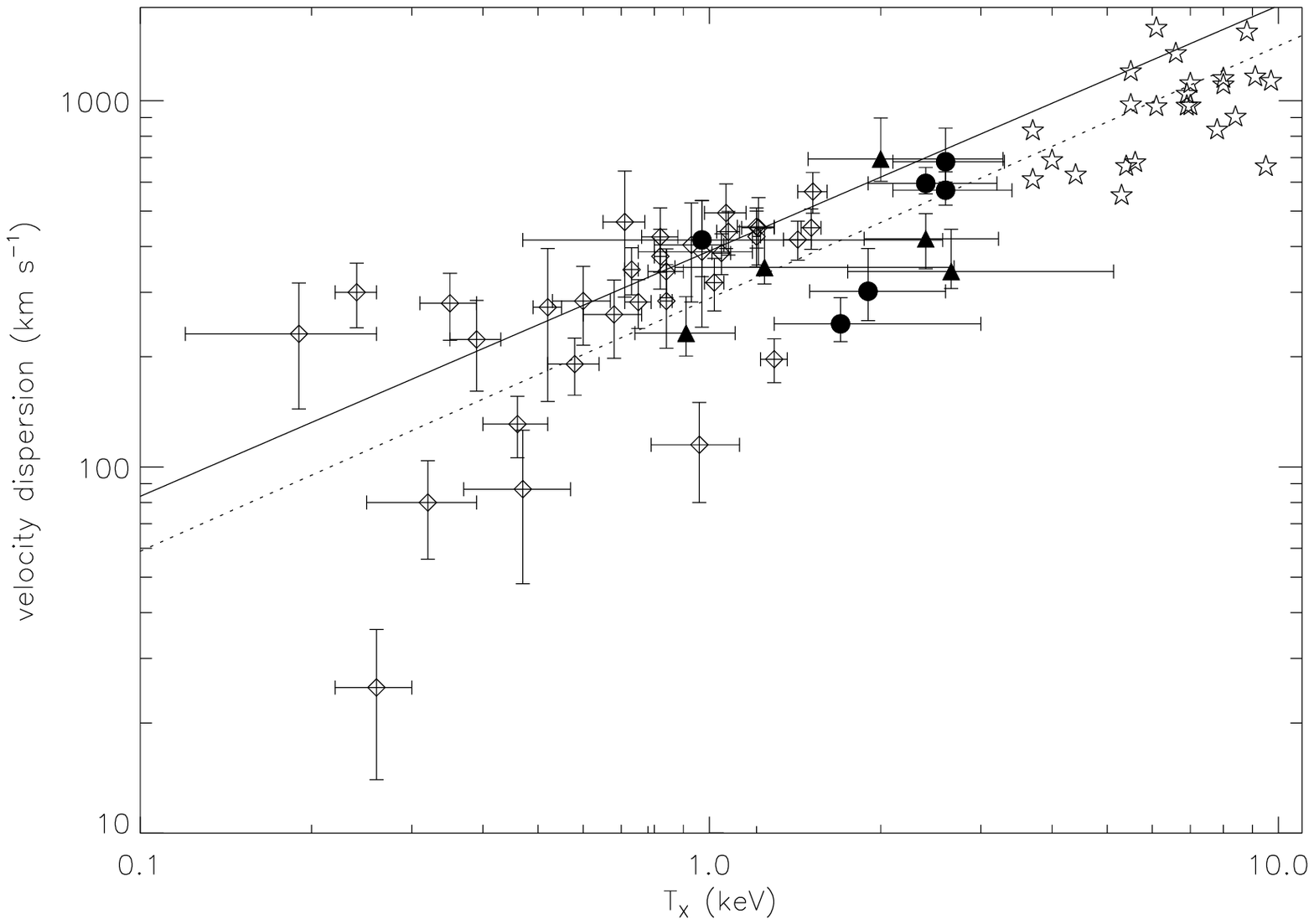}
\caption{ Top: $L_X-\sigma$ relation for our groups (filled circles), the low-redshift GEMS groups 
(Osmond \& Ponman 2004)(open diamonds), the cluster sample of Markevitch (1998) (open stars), and the 
moderate-redshift XMM-LSS groups (filled triangles)(Willis et al. 2005).  Velocity dispersions for the 
Markevitch sample are taken from Horner (2001).  Also shown is the fit to the GEMS groups (solid line) and 
a fit to the Markevitch clusters (dotted line)(Helsdon \& Ponman, in preparation).  RXJ1347+07, for which 
we do not have XMM data, is plotted with a solid square using the ROSAT luminosity.  Bottom: Same for the 
$\sigma-T$ relation.  The error bars on our points show the 90\% uncertainty in temperature, the 1$\sigma$ 
errors in $\sigma$, and the error in luminosity from both the uncertainty in temperature and metallicity.  
Error bars for the other samples are 1$\sigma$. }
\end{figure}
\clearpage
Previous studies have compared the ratio of the specific energy in the motion of the group galaxies to that in 
the X-ray emitting gas through the parameter $\beta_{spec} = \mu m_p \sigma_v^2/kT_X$, where $\sigma_v$ is the
line-of-sight velocity dispersion of the group galaxies, $T_X$ is the X-ray temperature, $k$ is the Boltzmann
constant, and $\mu m_p$ is the mean particle mass in the gas.  In Figure 3, we plot $\beta_{spec}$ for the
six groups with XMM-Newton observations versus X-ray luminosity.  Also shown are the intermediate-redshift groups
studied by Willis et al. (2005) as part of the XMM-LSS survey.  Most of our groups are consistent with 
$\beta_{spec} = 1$, but two groups, RXJ0329+02 and RXJ1334+37, fall well below this number.  Including the 
Willis et al. (2005) groups roughly half of intermediate-redshift groups appear to have $\beta_{spec} < 1$.
This deviation from $\beta_{spec} = 1$ is not seen in groups at low-redshift at least at these luminosities
(Mulchaey \& Zabludoff 1998; Osmond \& Ponman 2004), although some very low velocity dispersion systems may
have $\beta_{spec} < 1$ (Osmond \& Ponman 2004).  In RXJ1334+37 we have fairly good membership with 17 galaxies,
but in this system both the offset of the BGG from the X-ray peak and the possibility of 
substructure discussed
in the previous section indicate that this group may not be dynamically relaxed.  The other groups with 
$\beta_{spec} < 1$ have on the order of 10 known members.  In addition, there is significant scatter in the 
relationship between velocity dispersion and temperature for these groups.  
A more detailed understanding of the dynamics of these systems is needed to understand and 
confirm the deviation from equipartition between the galaxies and gas seen here.  We do not observe a correlation 
of $\beta_{spec}$ with luminosity over this limited luminosity range.

It should be noted that for the Markevitch (1998) clusters shown in Figure 2 the X-ray properties have been 
corrected for the presence of cool cores, although in general the corrections are small.  We cannot constrain the
presence of a cool gas core in our groups, because the XMM-Newton data are not of sufficient depth to construct temperature
profiles, so no correction has been applied to the intermediate-redshift groups (our sample or the Willis et al. 2005
sample).  The generally large core radii of our groups (see Paper II) indicate that they do not contain a strong
surface brightness peak.  If a cool core was present, excluding this component would decrease the luminosities but
increase the temperatures, thus decreasing $\beta_{spec}$.  Osmond \& Ponman (2004) looked for the presence of cooler
gas in the low-redshift GEMS groups with sufficient statistics to construct temperature profiles and found that
the average temperature drop in the group centers was only 4\% and the corrections for this component were small.
\clearpage
\begin{figure}
\epsscale{0.8}
\plotone{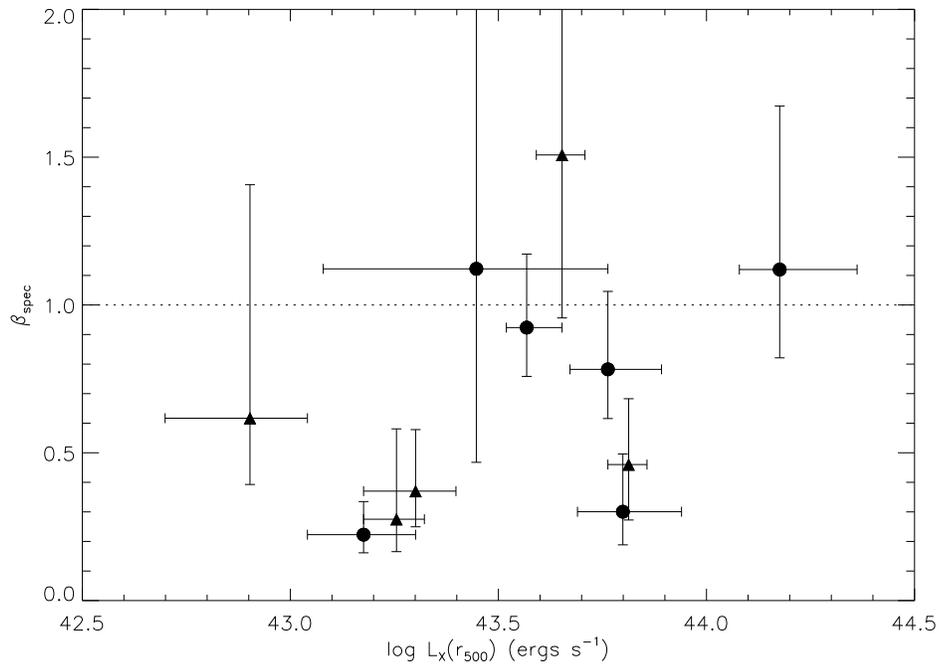}
\caption{ Values of $\beta_{spec}$ versus X-ray luminosity for the six groups observed with XMM (filled circles).  
Also shown are the intermediate-redshift XMM-LSS groups (filled triangles)(Willis et al. 2005). }
\end{figure}
\clearpage

\section{ MORPHOLOGICAL COMPOSITION }

\subsection{ Early-type Fraction }

At low redshifts, groups with significant X-ray emission are found to contain significant numbers of early-type 
galaxies, while spiral-only groups tend to be lower-mass systems (Mulchaey et al. 2003; Osmond \& Ponman 2004).
  In X-ray emitting groups, the 
fraction of early-type galaxies ($f_e$) can be as high as those found in clusters (Zabludoff \& Mulchaey 1998).  
Table 3 lists the fraction of 
early-type galaxies (E and S0) for the seven groups discussed here.  Where possible we use the HST morphologies, 
but for the five groups with Gemini imaging we use the Gemini morphologies for galaxies that fall outside of the 
HST field of view.  We also list the early-type fractions after limiting the sample to galaxies with radii less than 
700 $h_{70}^{-1}$ kpc and brighter than $M_V=-20.5 + 5$log$_{10} h_{70}$ where we are fairly complete.  These restrictions turn out to have little 
effect on the early-type fractions.  Limiting the sample to only those galaxies with HST morphologies also has little
effect but further limits the statistics, and here we consider only the total early-type fraction for all groups.
For galaxies with HST morphologies meeting our completeness limits ($M_V<-20.5 + 5$log$_{10} h_{70}$ and $R<700$ 
$h_{70}^{-1}$ kpc) and combining the 
data for all seven groups, we find that 
intermediate-redshift groups are composed of 68$\pm8$\% early-type galaxies (for 37 galaxies).  
Considering also galaxies with Gemini morphologies meeting the completeness limits yields a similar fraction 
of 70$^{+7}_{-5}$\% (for 57 galaxies).  Errors are determined through bootstrap resampling.

We would like to compare the morphological composition of our groups to similar groups at low redshift.  We choose
as a comparison sample the five X-ray luminous groups in Zabludoff \& Mulchaey (2000) (NGC 2563, HGC 42, NGC 4325, 
HGC 62, and NGC 5129).  In addition to being X-ray luminous, massive groups, these groups have the most complete
spectroscopy, sampled out to similar radii as our groups, of any low-redshift group sample.  Imposing the same limits
of $M_V<-20.5 + 5$log$_{10} h_{70}$ and $R<700$ $h_{70}^{-1}$ kpc, we find an early-type fraction of 50$\pm13$\% 
(24 galaxies) in low-redshift groups.  This 
fraction is a bit lower than what we found for our groups; however, Zabludoff \& Mulchaey (1998) employ an intermediate
morphological category, S0/a.  If these galaxies are classified as early-types rather than late-types, we find 
$f_e=0.63^{+0.12}_{-0.09}$
in low-redshift groups.  Although this fraction is still a bit lower than ours, the difference is not significant 
given the number of galaxies sampled.

As found in Paper I based on more limited group membership, X-ray emitting groups at intermediate redshifts 
already contain a large fraction of early-type galaxies similar to the numbers found in low-redshift, X-ray luminous groups.
These large early-type fractions are similar to those found in clusters (Dressler et al. 1997; Treu et al. 2003).

The field, however, is dominated by spiral galaxies (e.g. Oemler 1992; Smail et al. 1997).  Considering all
of the galaxies in our sample with redshifts in the same range as our groups ($0.2<z<0.6$) that are not group
members, we define a field sample selected and observed in the same manner as our group galaxies.  These galaxies
have a similar distribution in magnitude and redshift as the group members.  The field 
early-type fraction shows more sensitivity to the use of Gemini morphologies, most likely because the larger
fraction of late-type galaxies leads to a greater sensitivity to the misclassification of spirals as early-types
from the ground-based imaging, so we consider only those galaxies with HST imaging.  We find a field early-type
fraction of 19$^{+6}_{-5}$\% (36 galaxies) for all galaxies and of 29$^{+12}_{-11}$\% (17 galaxies) for galaxies 
brighter than $M_V=-20.5 + 5$log$_{10} h_{70}$.
The limitation of using galaxies with HST morphologies limits our field sample size, but these fractions are 
similar to those found by other studies (Dressler et al. 1997; Treu et al. 2003).  

As found in previous studies, we find that groups have significantly higher fractions of early-type galaxies 
than the field.  The fact that
our massive groups generally have early-type fractions as high as those found in clusters supports the idea that
the morphological transformation of galaxies occurs in lower density environments before they are incorporated into 
massive clusters.  Similarly, Balogh et al. (2002b) find a small fraction of disk-dominated galaxies ($B/T<0.4$)
in $z\sim0.25$, low-luminosity clusters.

Although we find an overall early-type fraction of $\sim 70\%$, the individual groups in our sample show a large
range of $f_e$ values.  In particular, both RXJ1205+44 and RXJ1347+07 have $f_e \sim 0.4$ for galaxies brighter than 
$M_V=-20.5 + 5$log$_{10} h_{70}$, nearly as low as the field.  The significance of this result is limited somewhat by the relatively 
small number of galaxies sampled per group.  However, assuming a true group early-type fraction of 70\%, we can 
calculate the probability of finding as few as $k$ early-type galaxies in $n$ Bernoulli trials.  For RXJ1205+44,
this yields a probability of 0.06 of finding only three early-type galaxies in a sample of eight.  This result is 
intriguing given that RXJ1205+44 is the most luminous, highest velocity dispersion group in our sample; it also 
has a fairly symmetric X-ray morphology with the X-ray emission centered on the BGG.  The low
$f_e$ is less significant for RXJ1347+07 with a probability of 0.13.  Using instead a bootstrap resampling of
the group galaxies to test the significance of these early-type fractions yields probabilities of 0.04 for 
RXJ1205+44 and 0.14 for RXJ1347+07 that the true early-type fractions are 70\% or higher (for 1000 bootstrap trials).

In Paper I, one other group, RXJ0210-39, was found to have a low early-type fraction, $f_e \sim 0.3$.  This group 
is also distinctive in that the most luminous galaxies near the group center are all spirals.  The nature of this 
group will be explored in a future paper with scheduled X-ray and deeper spectroscopic observations.

\subsection{ Distribution of Morphologies }

A number of studies have shown a correlation between galaxy morphology and radius in both clusters (e.g. Dressler 1980;
Dressler et al. 1997; Treu et al. 2003; Postman et al. 2005) and in low-redshift groups (Brough et al. 2006; 
Helsdon \& Ponman 2003).  In Figure 4
we show the relationship between early-type fraction and radius for a composite of our groups.  To achieve reasonable 
statistics,
we limit this study to three radial bins with at least 30 galaxies per bin.  Points are plotted at the outer radius
of each bin, and the errors were determined through bootstrap simulations.  Within a radius of at least 375 $h_{70}^{-1}$ kpc, the
early-type fraction is very high, $82\pm6$\% and $86^{+5}_{-6}$\% for the two inner bins, respectively.  Outside of 
this radius $f_e$ drops significantly to 68$\pm7$\% while still 
remaining significantly higher than the field.  This drop from the central early-type fraction is significant at 
better than 95\% confidence.  The outer bin includes galaxies with radii of nearly 1.2 $h_{70}^{-1}$ Mpc, but 
limiting the outer radius to 700 $h_{70}^{-1}$ kpc, the radius sampled in all groups, does not change the results.  Most of the
galaxies in this bin are within 700 $h_{70}^{-1}$ kpc, and their median radius is 560 $h_{70}^{-1}$ kpc.  Increased statistics and 
wider field spectroscopy are needed to establish at what radii the decline in the number of early-type galaxies 
occurs and at what radius it eventually reaches the field value.  However, the fraction of early-type galaxies remains
higher than the field out to at least $r_{500}$, where $r_{500} \sim 600$ $h_{70}^{-1}$ kpc for these groups (Paper II).
A similar trend of early-type fraction with radius is seen in clusters (Dressler et al. 1997; Treu et al. 2003) 
and in groups at low redshift (Brough et al. 2006; Helsdon \& Ponman 2003).

Galaxy morphology is also found to correlate strongly with local galaxy density (e.g. Dressler 1980;
Dressler et al. 1997; Treu et al. 2003).  The low galaxy densities in these groups mean that errors in the 
background subtraction dominate outside the group cores, making a study of the morphology density relation
unfeasible.  We note, however, that within a radius of 375 $h_{70}^{-1}$ kpc the average density of our groups is 17
galaxies $h_{70}^{2}$ Mpc$^{-2}$, and in the annulus between 375 $h_{70}^{-1}$ kpc and 700 $h_{70}^{-1}$ kpc this density 
is 1 galaxy $h_{70}^{2}$ Mpc$^{-2}$.
These numbers were derived to the same limiting magnitude used by other studies ($M_V = -20.4$ for $H_0 = 50$
km s$^{-1}$ Mpc$^{-1}$; Dressler et al. 1997) and using the background number counts from Postman et al. (1998).
Considering instead the two lowest-redshift groups, RXJ0720+71 and RXJ1256+25, for which we have nearly complete
spectroscopy at all radii and utilizing the spectroscopically confirmed members the average galaxy densities
are similar at 14 galaxies $h_{70}^{2}$ Mpc$^{-2}$ ($R \leq 375$ $h_{70}^{-1}$ kpc) and 3.7 galaxies $h_{70}^{2}$ 
Mpc$^{-2}$ ($375 < R \leq 700$ $h_{70}^{-1}$ kpc).

At similar densities, the early-type fractions in intermediate-redshift and local clusters are significantly lower
($f_e \sim 0.4-0.6$ for the density of our inner bin)(Dressler et al. 1997; Treu et al. 2003).  
A similar behavior was found by Helsdon \& Ponman (2003) when comparing 
low-redshift groups and clusters.  Helsdon \& Ponman (2003) suggest that this offset could be due to the higher merger
rate in groups.  This offset is intriguing as it may imply that galaxy transformation processes are efficient at lower
densities in groups than in clusters, but some care should be taken in interpreting these results as there are a number of differences
between the cluster and group studies mentioned here.  The cluster
morphology density relation is based on photometry, the magnitude limits of the various samples are not the same 
(but within a magnitude), and the group studies employ average galaxy density rather than local density.
\clearpage
\begin{figure}
\plotone{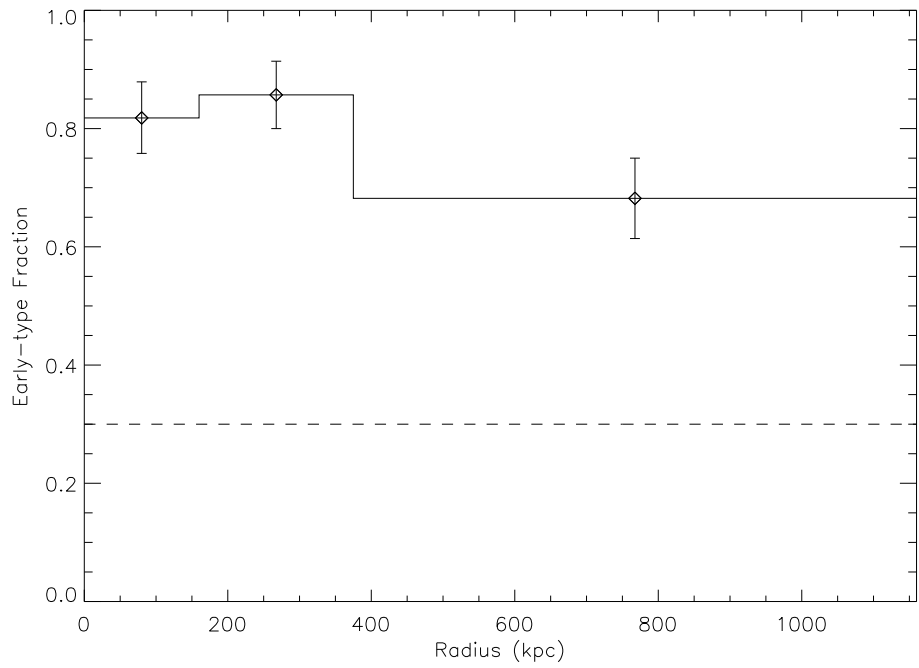}
\caption{ Early-type fraction versus radius.  Points are plotted at the central radius of each bin. Error
bars are derived through bootstrap resampling.  The dashed line shows the field early-type fraction. }
\end{figure}
\clearpage
Dressler et al. (1999) also find a difference in the velocity dispersions of different morphological types in their
intermediate-redshift MORPHS clusters.  They find that elliptical galaxies have a lower velocity dispersion than later
types.  We do not find a significant difference in the velocity dispersions of early and late-type galaxies in
intermediate-redshift groups.  These two classes have $\sigma_v = 500^{+41}_{-35}$ km s$^{-1}$ and 
$461^{+103}_{-56}$ km s$^{-1}$, respectively.  These numbers are quite similar and indistinguishable within the errors.
We note, however, that we have combined E and S0 galaxies in the early-type classification due to the limited
number of galaxies for which we have HST morphologies and the difficulty of distinguishing these two classifications.
Dressler et al. (1999) find that like the later types, S0 galaxies have a larger velocity dispersion than E galaxies.  Here we 
have simply used galaxy velocities relative to the central velocity of their respective groups, but the results do not 
change if we instead weight the velocities by group velocity dispersion.

\subsection{ Mergers/Interactions }

We also investigate the existence of mergers and interactions among the group galaxies.  These galaxies were classified
based on having multiple nuclei within a common envelope, tidal features, or signs of interaction with a close companion.
Besides the three BGGs with multiple nuclei (Paper I, see \S6), we find four group galaxies that show signs of a 
merger or interaction, one
in RXJ1256+25 imaged with Gemini and three in RXJ1648+60 imaged with HST (see Figure 5).  
This corresponds to roughly 5\% of the sample.
Including the BGGs increases this number to $\sim$ 10\%.  Three of the mergers are quite luminous with magnitudes brighter
than $M_V=-21.4 + 5$log$_{10} h_{70}$.  One is an Sb galaxy undergoing a minor merger, but the other three show two relatively 
bright components.  None of these galaxies, including the three BGGs, show significant [OII] emission.  
Similar bright mergers
lacking [OII] emission were observed in the high-redshift cluster MS1054-03 (van Dokkum et al. 2000; Tran et al. 2005b),
although in a higher fraction than we observe in our groups.
These mergers, particularly the BGGs, may be examples of the build-up of the brightest cluster/group galaxies through 
``dry mergers'' that do not induce significant star-formation.  Tran et al. (2005b) argue based on the mergers observed
in MS1054-03, that at least in the case of this cluster it is possible that all early-type members evolved from passive
galaxy-galaxy mergers at redshifts less than 1.  They estimate that for every present day cluster early-type galaxy to 
have undergone a merger at $z\leq1$ the average merger fraction would be $\sim6.5$\%.  This percentage is comparable to
the fraction of mergers we observe in intermediate-redshift groups; however, this is based on a small number of
merging galaxies.
\clearpage
\begin{figure}
\epsscale{0.4}
\plotone{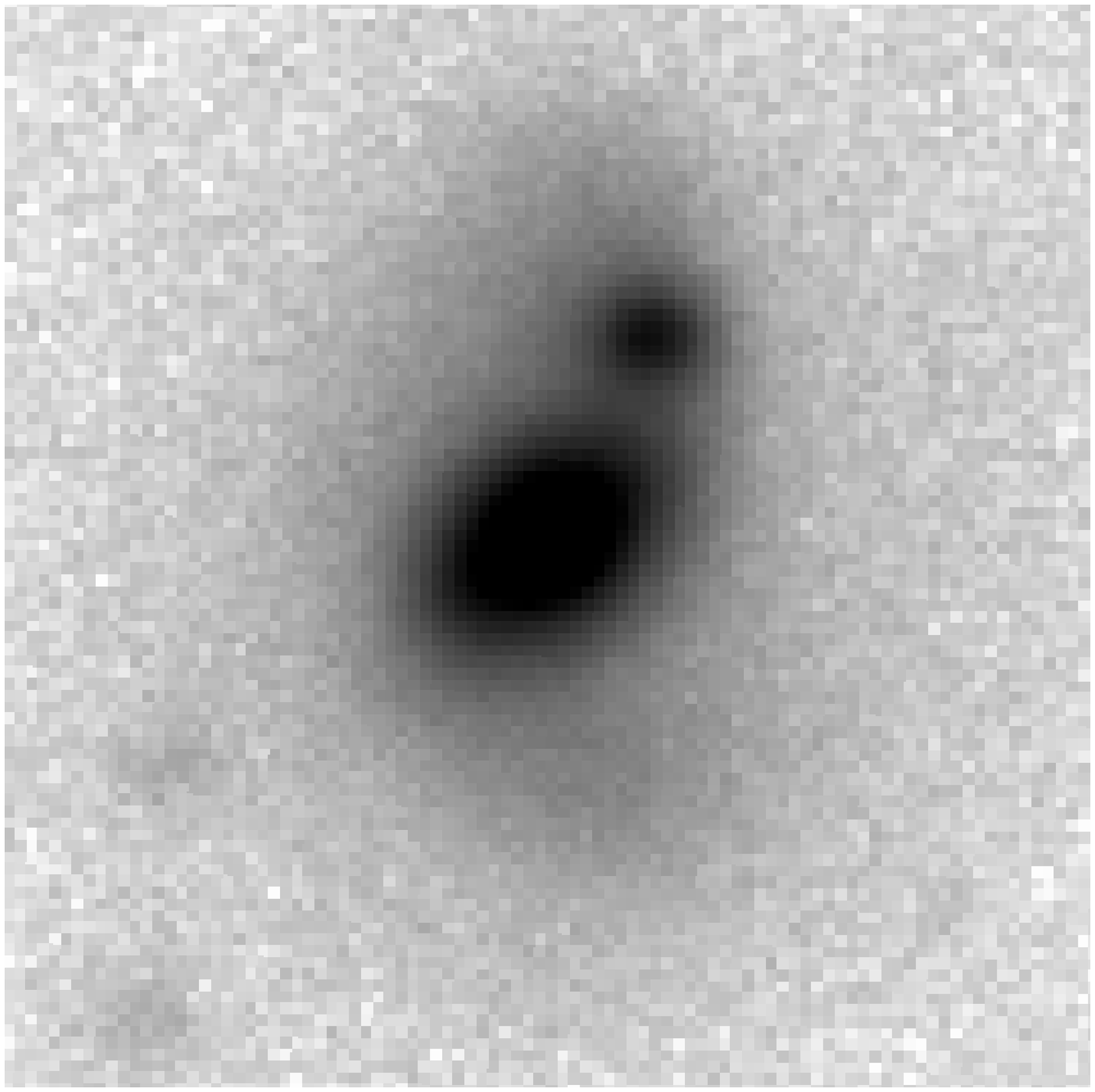}
\plotone{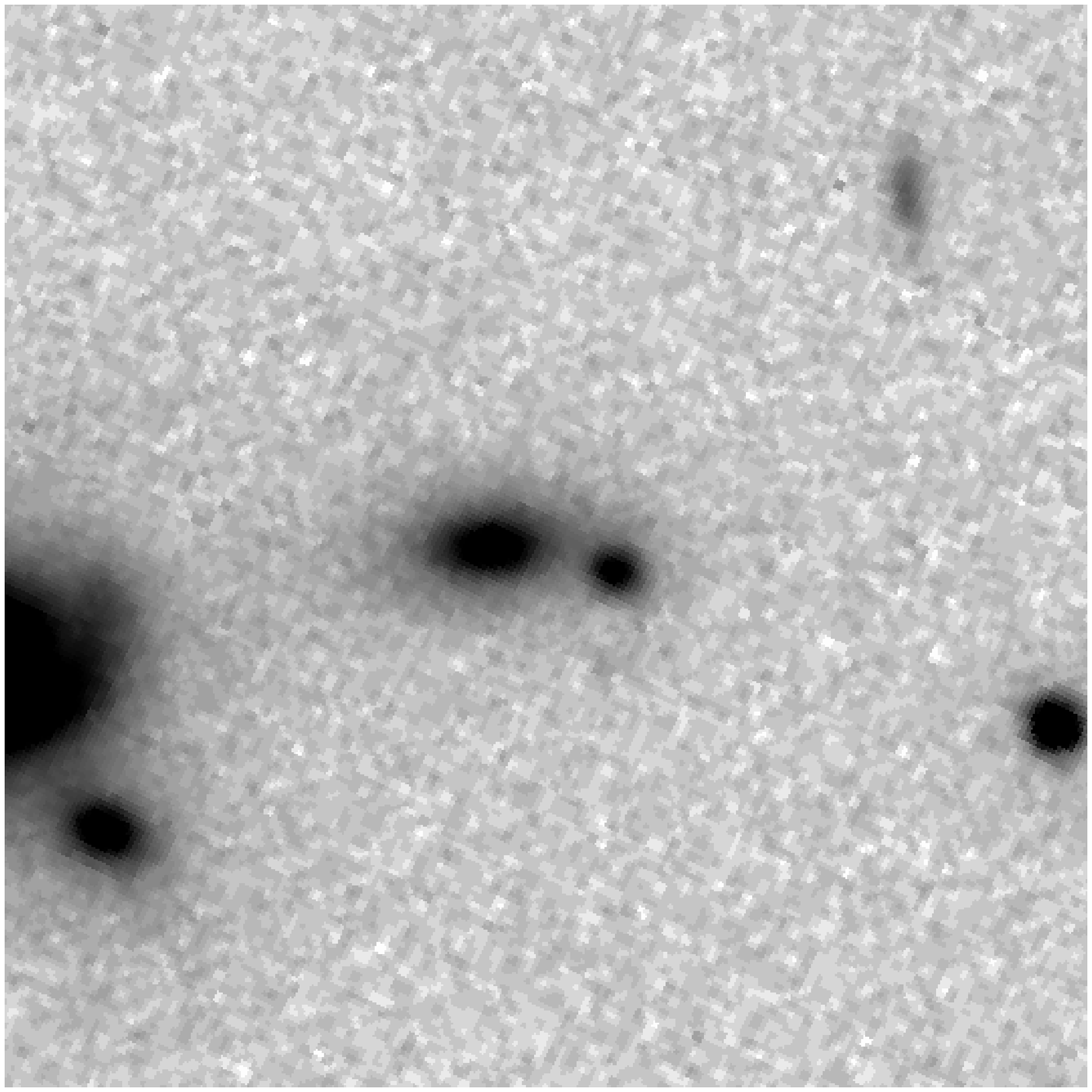}
\plotone{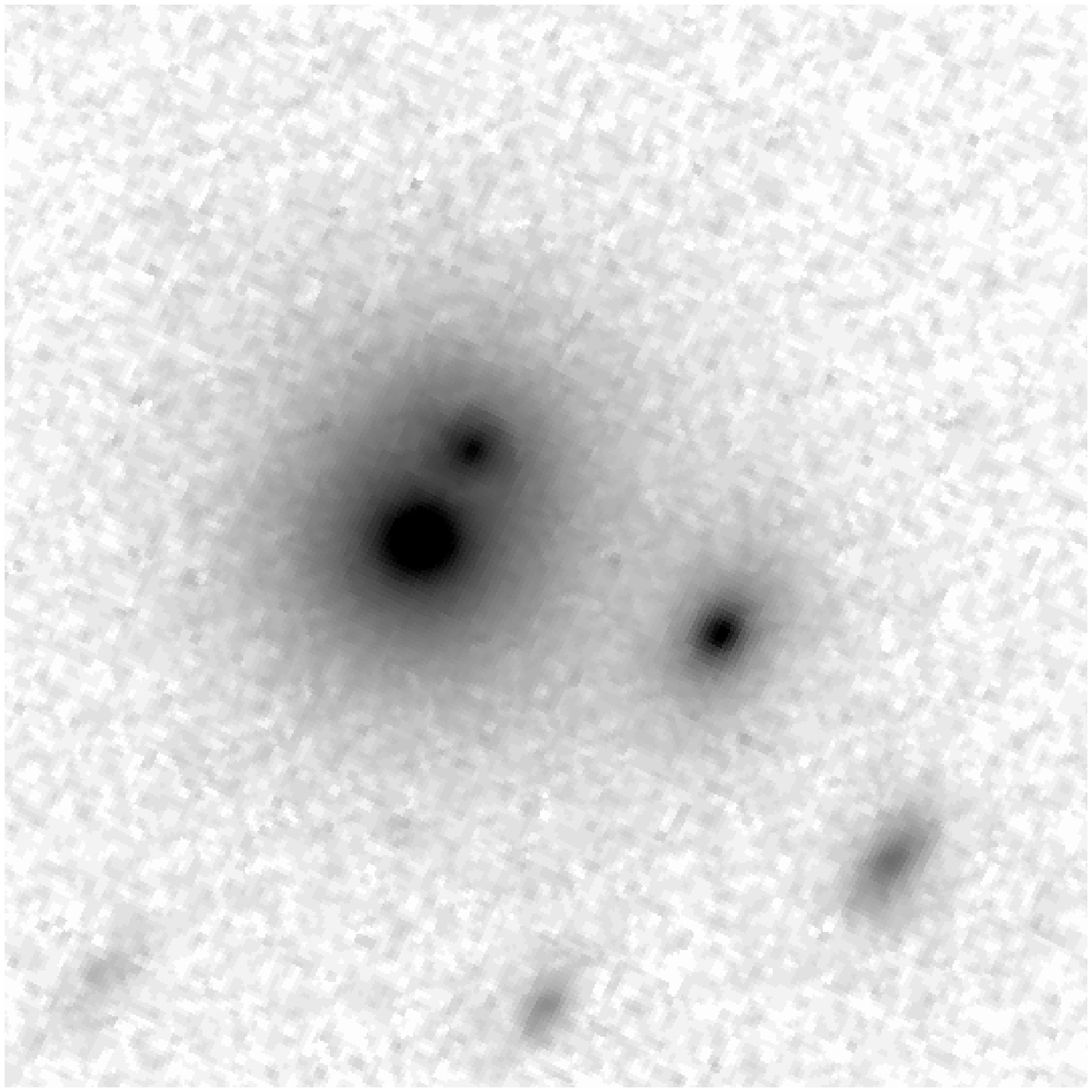}
\plotone{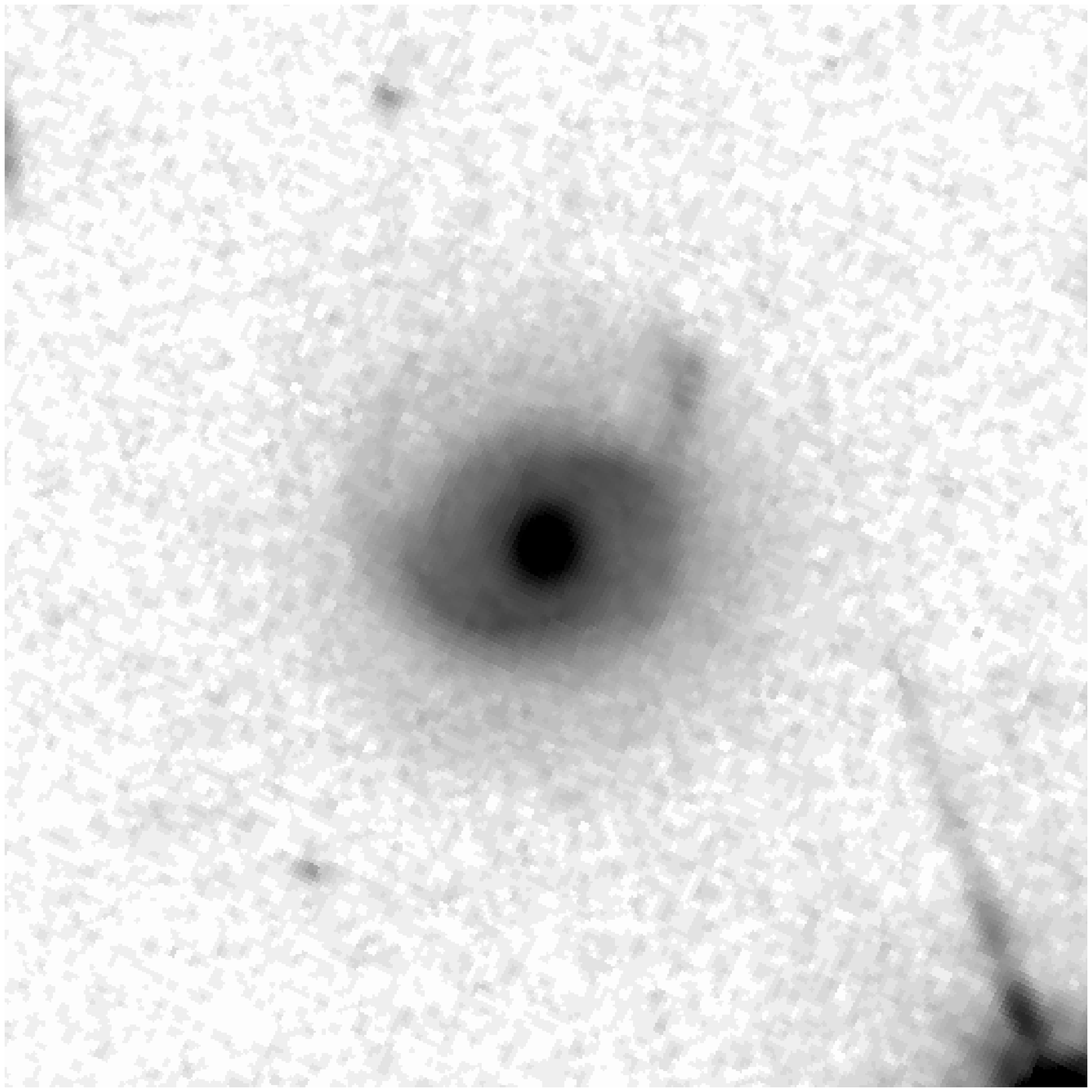}
\caption{ Images of the mergers/interactions observed among the group member galaxies.  The top left image is an r-band image
from Gemini, while the other three are F702W band HST images.  Images are approximately 50 $h_{70}^{-1}$ kpc on a side. }
\end{figure}
\clearpage

\section{ SPECTRAL PROPERTIES }

\subsection{ [OII] fraction }

We also investigate the fraction of group galaxies that show significant [OII]$\lambda$3727 emission, 
an indicator of active star-formation.  Similar to other high-redshift surveys, H$\alpha$ does not fall 
in the spectral range covered by our observations.  However, [OII] has been shown to correlate with H$\alpha$
emission (Hopkins et al. 2003; Jansen et al. 2000; Kenicutt 1992), although with large scatter, and Hopkins
et al. (2003) found that for SDSS galaxies the star-formation rates derived from [OII] emission are consistent
with those from H$\alpha$.  Here we are concerned only with selecting star-forming galaxies, and we do not 
attempt to calculate precise star-formation rates.  One further concern is the possible contribution of active galactic nuclei (AGNs) or
LINERs to these [OII]-selected galaxies; Yan et al. (2006) find that this could be important for
red galaxies.  Due to the redshift range and limited spectral coverage of our sample, we are only sensitive to 
[OII]$\lambda$3727, H$\beta$, and [OIII]$\lambda$5007 in a fraction of our galaxies.  For those galaxies with
[OII] emission and where both H$\beta$ and [OIII]$\lambda$5007 fall within the spectrum, we look for galaxies
with line ratios similar to AGNs ([OIII]$\lambda$5007/H$\beta > 3$; e.g., Kauffmann et al. 2003).  This is a 
rough demarcation since we do not have the additional constraint of [NII]$\lambda$6583/H$\alpha$.
We do not find any group galaxies with line ratios similar to AGNs, but we do find a few among the field galaxies.
Of the 29 field galaxies with significant [OII] emission and brighter than $M_V=-20.5 + 5$log$_{10} h_{70}$, we are sensitive to 
H$\beta$ and [OIII]$\lambda$5007 in eight; two of these have [OIII]$\lambda$5007/H$\beta > 3$, indicating that
they could be AGNs.  We conclude that AGNs do not contribute significantly to our group sample, but that they
may make some contribution to the field sample.  We do not attempt to correct for this possible AGN component,
because the small number of galaxies sampled makes this correction quite uncertain.  In addition, many galaxies
hosting AGNs also have significant star-formation (e.g., Cid Fernandes et. al 2004; Heckman et al. 1995).

We measure [OII] equivalent widths using two methods.  The first is the interactive fitting technique used by Dressler
et al. (1999; 2004). The second is a typical bandpass technique where the continuum is determined through a 
linear fit to red and blue bandpasses and then the flux above this continuum is summed in the line bandpass.  
Here we use the bandpass
definitions of Balogh et al. (1999).  These two methods give similar equivalent widths (typically within 15\%),
particularly for galaxies with significant [OII] emission.  In this study, we consider only the groups observed 
with Gemini as the Keck spectra are typically of much poorer quality.

Following previous studies (Wilman et al. 2005b; Zabludoff \& Mulchaey 1998; Dressler et al. 1999), we define 
galaxies with significant [OII] emission as
those with EW[OII]$>$5\AA.  Table 3 lists the fraction of galaxies with significant [OII] emission ($f_{\rm [OII]}$) 
for each group for both all of the group members with Gemini spectra and only those galaxies with 
$M_V<-20.5 + 5$log$_{10} h_{70}$ and $R<700$ $h_{70}^{-1}$ kpc.  Here we list the results from the interactive fitting technique, but in only one 
case does the fraction change at all for the bandpass technique, and in this case the difference is one galaxy.

We find a total emission-line fraction in intermediate-redshift groups of $f_{\rm [OII]}=0.28^{+0.05}_{-0.06}$ (76 galaxies) 
or $f_{\rm [OII]}=0.30^{+0.05}_{-0.10}$
(40 galaxies) considering only those galaxies meeting our completeness limits.  We find a similar fraction of
emission-line galaxies in our comparison sample of low-redshift, X-ray emitting groups (Zabludoff \& Mulchaey 2000).  
Here $f_{\rm [OII]}=0.21\pm0.08$ 
(24 galaxies) for $M_V<-20.5 + 5$log$_{10} h_{70}$ and $R<700$ $h_{70}^{-1}$ kpc.  This fraction is a bit lower than our intermediate-redshift fraction,
but not at high significance (probability of 0.11).  In contrast, the same fractions 
for the intermediate-redshift field are $f_{\rm [OII]}=0.59^{+0.04}_{-0.05}$ (112 galaxies) and $f_{\rm [OII]}=0.49\pm0.07$ 
(59 galaxies) for galaxies brighter than $M_V=-20.5 + 5$log$_{10} h_{70}$.

As with the fraction of early-type galaxies, we find that X-ray emitting groups contain a similar fraction of
emission-line galaxies to massive clusters.  Poggianti et al. (1999) find that 32\%
of the galaxies in the intermediate-redshift MORPHS cluster sample show significant [OII] emission, and Balogh
et al. (1999) find a fraction of 25\% for the intermediate-redshift CNOC1 clusters.  Due to weightings applied 
to these samples to correct for sample selection and incompleteness we cannot reproduce our exact magnitude 
limits, but the cluster samples are of roughly similar depth to ours (within less than a magnitude) 
even after their completeness corrections.  As found for clusters, there is a significant deficit of
galaxies with [OII] emission in groups relative to the field, indicating a deficit of star formation.  Balogh et
al. (2002b) studied a sample of low-luminosity, X-ray selected clusters at $z\sim0.25$ and found that these
systems had low [OII] fractions (22\%) similar to intermediate-redshift massive clusters.  Our study extends these
results to both higher redshift and lower mass.  

This result is seemingly at odds with Poggianti et al. (2006),
who find a correlation between [OII] fraction and velocity dispersion for the optically-selected EDisCS clusters 
and groups at $0.4 \leq z \leq 0.8$, but there are several differences between these studies.  The EDisCS systems are 
generally at higher redshift, are sampled to slightly larger radii ($r_{200}$ vs. $\sim 1.2 r_{500}$), and a lower
cut-off of EW[OII]$>$3\AA\ was used in this study.  In addition, optically-selected groups tend to be less X-ray
luminous, possibly indicating that they are less dense/relaxed than the systems studied here, which could lead to higher
[OII] fractions for the group scale systems in the Poggianti et al. (2006) sample.  However, 
considering only those systems in Poggianti et al. (2006) with redshifts and velocity dispersions in the same ranges
as our systems, we find an average [OII] fraction for their groups of $f_{\rm [OII]}=0.26$ (compared to 0.36 for our sample 
if we use EW[OII]$>$3\AA), very similar to our X-ray selected systems.  The groups with high [OII]
fractions in the EDisCS sample seem to be at higher redshifts.  Wilman et al. (2005a,b) study star-formation in a sample 
of optically-selected, intermediate-redshift ($0.3 < z \leq 0.55$) groups from the CNOC2 survey.  These 
systems seem to have higher [OII] fractions than found here for X-ray emitting
groups, but a direct comparison is not possible due to differences in selection and magnitude limits.

We do not see a significant evolution in fraction of emission-line galaxies compared to X-ray selected, low-redshift 
groups, although a small evolution could be masked by the small numbers.  Similarly, Nakata et al. (2005) find that 
the fraction of galaxies with strong [OII] emission (EW[OII]$>$10\AA) in the central 700 $h_{70}^{-1}$ kpc of clusters 
does not evolve significantly with redshift, although this study probes both stronger [OII] and denser regions.  
In contrast, they find a strong increase in the number of field galaxies showing 
significant [OII] emission with increasing redshift.
Wilman et al. (2005b), on the other hand, find that $f_{\rm [OII]}$ increases with redshift both in the field and in 
groups, but with $f_{\rm [OII]}$ always smaller in groups than in the field at a given epoch.  This study considers 
optically-selected groups from the CNOC2 and 2dF surveys which are generally lower mass systems than our X-ray 
selected groups.  The
difference in evolution could then indicate the mass/density regime in which the truncation of star formation
becomes important.  Poggianti et al. (2006) also find a significant evolution in [OII] fraction with redshift, 
but here the difference is most pronounced for clusters.  Their low-redshift groups, selected optically from
the SDSS, show significant scatter in $f_{\rm [OII]}$.  After combining all of the low-redshift systems in their sample 
in our velocity dispersion range we find $f_{\rm [OII]}=0.32$ which is very similar to the average for both our groups and 
for the intermediate-redshift EDisCS groups.  A number of factors could influence these results, and larger, deeper 
samples are needed to 
determine at what halo mass, redshift, local density, or radius the evolution of star-formation occurs.  It
should also be noted that while it is the best comparison sample available of well-studied, X-ray luminous 
groups, the groups in our low-redshift sample tend to have slightly lower masses than our intermediate-redshift 
groups, which could partially mask an evolution in star formation.
 
As with $f_e$, we find that $f_{\rm [OII]}$ varies among the groups.  In particular in RXJ1347+07 67$^{+16}_{-17}$\% of the 
group galaxies show significant [OII] emission, a higher fraction than was found here for the field.  Again 
assuming a binomial distribution and a true fraction of 30\%, the probability of finding an $f_{\rm [OII]}$ as high
as or higher than 67\% for this group is 0.07 considering only the bright galaxies or 0.03 considering all galaxies.
Using a bootstrap resampling of the group galaxies, these probabilities are 0.01 and 0.004, respectively.  RXJ1347+07
has the lowest velocity dispersion of any group in our sample, and its young galaxy population could possibly be
because it is a lower mass system.  However, RXJ1205+44 which contains a high fraction of spiral galaxies has the
highest velocity dispersion in the sample.
Unfortunately, the small number of galaxies per group leads to relatively large errors on the galaxy populations of 
individual groups and precludes a study of the correlation of $f_e$ and $f_{\rm [OII]}$ with redshift or velocity 
dispersion, but we do observe a significant variation in these properties in systems that all have significant X-ray
emission.  The small numbers here highlight the difficulty of studying these relatively low-density environments.

We also calculated H$\delta$ equivalent
widths for our group galaxies, but this measurement proved to be much more sensitive to the method used.  
We therefore delay discussion of the more detailed spectral types of group galaxies until a future paper 
where the measurement of Balmer absorption will be investigated in greater depth.  We note, however, that
while the fraction of these galaxies depends on the chosen method of measuring equivalent widths, we do
find at least a few group members with post-starburst like spectra (EW[OII]$<5$ \AA\ and H$\delta > 3$ \AA).

\subsection{ Distribution of [OII] Emitters }

We also investigate the dependence of the fraction of emission-line galaxies on radial distance from the group center, shown in Figure 6.
Here we limit our study to simply small ($R \leq 375$ $h_{70}^{-1}$ kpc) and large radii ($R > 375$ $h_{70}^{-1}$ kpc), with each bin containing
38 galaxies.  We find a very significant increase (better than 99.9\% confidence based on bootstrap simulations) in
the fraction of galaxies with significant [OII] emission.  At small radii this fraction is $0.16\pm0.05$, and at 
large radii it is $0.40^{+0.07}_{-0.08}$.  However, as we found for the fraction of early-type galaxies, even in the 
outer bin the [OII] fraction is
significantly lower than the field value.  The median radius of this outer bin is 580 $h_{70}^{-1}$ kpc, and our results 
do not change
if we limit the sample to galaxies at $R<700$ $h_{70}^{-1}$ kpc.  Here we observe a relatively strong evolution in [OII] 
fraction with
radius, but this fraction remains lower than the field out to at least $r_{500}$.  A similar, although possibly weaker,
 trend of increasing [OII] fraction with radius has also been observed in intermediate-redshift clusters 
(Balogh et al. 1999) and optically-detected, intermediate-redshift groups (Wilman et al. 2005b).
\clearpage
\begin{figure}
\epsscale{0.8}
\plotone{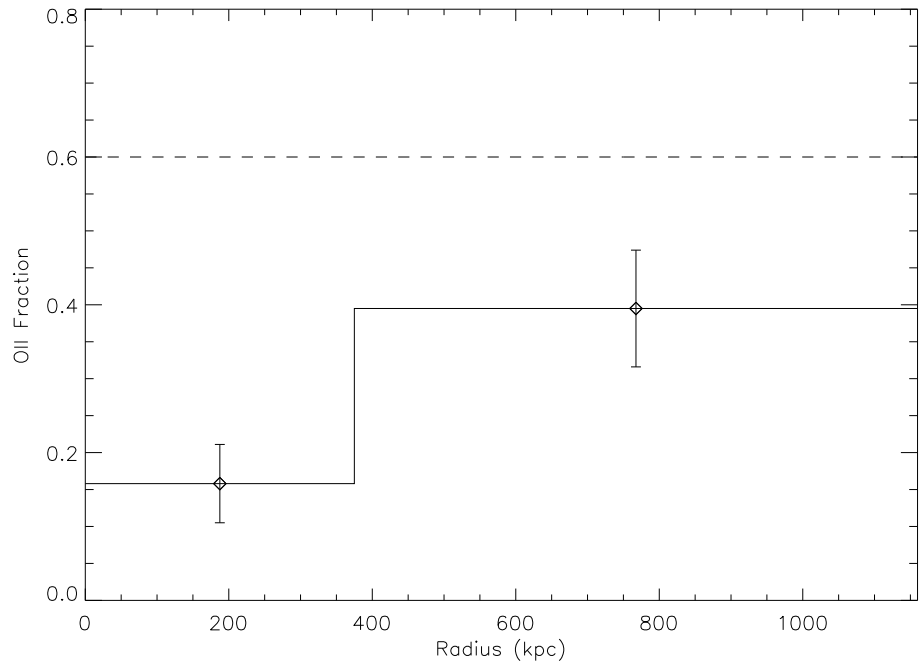}
\caption{ [OII] fraction versus radius.  Points are plotted at the central radius of each bin. Error
bars are derived through bootstrap resampling.  The dashed line shows the field [OII] fraction. }
\end{figure}
\clearpage
Despite this strong radial gradient, we again do not find a significant difference in the velocity dispersion
of the galaxies with [OII] emission.  We find $\sigma_v = 431^{+116}_{-73}$ km s$^{-1}$ and 
$449^{+55}_{-32}$ km s$^{-1}$ for galaxies with and without [OII] emission, respectively.  
For the MORPHS clusters, Dressler et al. (1999) instead find that galaxies
with significant emission have a higher velocity dispersion than passive galaxies.  We cannot rule out small differences
in velocity dispersion; however, the best-fit velocity dispersions  for these two populations are quite similar.  
Balogh et al. (2002b) also do not find a significant difference in dispersion for active and passive galaxies
in their low-luminosity clusters.

\subsection{ Morphologies of [OII] Emitters }

We now consider the relationship between star-formation and galaxy morphology.  In general, galaxies with significant 
star formation are found to be late-types, and we find that the fraction of galaxies with [OII] emission is very similar
to the fraction of late-type galaxies.  However, some group ellipticals show significant [OII] emission, and a significant
number of group spirals do not show significant emission.  For the full sample, we find that $15^{+4}_{-6}$\% of 
early-type galaxies have [OII] emission, and $41 \pm 9$\% of spirals do not.  If we consider only galaxies with 
HST morphologies the sample of 
early-type galaxies is reduced to 19, none of which show [OII] emission.  We expect that a couple of the eight early-type 
galaxies (classified by the Gemini imaging) with significant [OII] emission may be misclassified spirals; however, 
all of these galaxies lie at radii greater than 350 $h_{70}^{-1}$ kpc where the [OII] fraction is significantly higher.
In comparison, Zabludoff \& Mulchaey (1998) find that
12\% of the elliptical galaxies in their low-redshift groups have star-forming or post-star-forming spectra, although 
both the magnitude limits and methods of classifying spectra differ from our sample.  We do not expect spirals to 
be misclassified in the ground-based imaging, and in fact the fraction of passive spirals is 46$^{+16}_{-15}$\% 
even when considering only spirals 
observed with HST.  The fraction of passive spirals in our intermediate-redshift groups is similar to the fraction
found in the $z\sim0.4$ cluster sample of Poggianti et al. (1999) and in the $z\sim0.25$ low-luminosity clusters of
Balogh et al. (2002b).
In contrast, in the field only $15 \pm 5$\% of spirals are passive (23$\pm9$\% for HST).  
Of the six passive group spirals 
imaged with HST, five of these were classified as early-type Sa or Sb spirals, and only one was classified as an Sd
galaxy.  The existence of passive spirals may be an indication that star-formation is shut off prior to the 
morphological transformation of galaxies in dense environments (Poggianti et al. 2006), and the existence of this
population in X-ray emitting groups and low-luminosity clusters could indicate that the suppression of star-formation
occurs in environments less massive than $\sigma \sim 400$ km s$^{-1}$.  However, a more detailed study of the 
spectra of passive spirals, and their correlation with environment is needed.

\section{ BRIGHTEST GROUP GALAXIES }

In Papers I and II, it was noted that at least one third of the groups in our sample do not have a dominant early-type
galaxy at the peak of the X-ray emission.  Specifically, two of the nine groups in Paper I contain a string of bright 
galaxies rather than one dominant galaxy (RXJ1648+60 and RXJ0210-39), and in RXJ1334+37 the BGG is offset by about
100 $h_{70}^{-1}$ kpc from the X-ray peak.  In addition, in three of the four groups where a bright elliptical is found at the center
of the X-ray emission the BGG has multiple components.  Two of these BGGs are the lowest-redshift groups in our sample, 
RXJ0720+71 and RXJ1256+25, which both show three luminous components within a common envelope. 
In contrast, similar examples, at least at these X-ray 
luminosities, are not seen at low-redshift (Paper I; Mulchaey et al. 2003; Osmond \& Ponman 2004).  
This difference is surprising given the general similarity between these
groups and X-ray emitting, low-redshift groups.  They have high X-ray luminosities and temperatures indicating a 
collapsed, dense system, and they generally follow scaling relations between $L_X$, $T_X$, and $\sigma_v$.  As shown in
this paper, they also generally have the high early-type fractions and low emission-line fractions typical of dense 
environments.  While there is significant variation in these properties from group to group, both RXJ1334+37 and
RXJ1648+60 have evolved galaxy populations.  In RXJ1334+37, there are possible indications of substructure (\S3.1). 
The current X-ray observation of RXJ1648+60, which was highly contaminated by flares, is insufficient to determine
its structure aside from the fact that the X-ray emission traces the group galaxies.  The other four groups observed
with XMM-Newton have symmetric, relaxed looking X-ray distributions with only RXJ0329+02 having significant ellipticity.

Table 4 lists the properties of the brightest galaxies in each group including distance from the X-ray center, 
offset from the mean group velocity, HST morphology, absolute magnitude, and the difference in magnitude between
the brightest and second brightest group members.  For RXJ1347+07, we do not have an XMM-Newton observation, so the group
center is taken to be the position of the BGG.  For RXJ0720+71 and RXJ1256+25, we list the BGG magnitude as the 
total magnitude of all three components and the difference between this total magnitude and the next brightest
group galaxy.  The magnitudes of the individual components in these BGGs differ by less than a magnitude.  
RXJ1256+25 was identified in the literature as a fossil group (Jones et al. 2003).  If its BGG is taken as a 
single component this galaxy would be significantly brighter than the next brightest member, but the difference in
magnitude is still only 1.5 mag.

All of the BGGs are elliptical galaxies, and with the exception of RXJ0329+02 they are all fairly round. In addition,
none of the BGGs show [OII] emission.  The BGGs 
in RXJ1334+37 and RXJ1648+60 are offset from the group center by $\sim100$ $h_{70}^{-1}$ kpc, but
this distance is less than $0.25r_{500}$.  Several of the BGGs are also offset from the mean group velocity, but in 
all cases this offset is less than $0.65\sigma_v$.  The typical errors on the central velocity are around 100 km
s$^{-1}$, and only the BGG in RXJ1347+07 has a velocity that could be significantly different from the 
central location ($\sim$ 95\% confidence).  Here the increased membership information and higher-S/N spectra 
significantly improve both the determination of the central velocity and the individual galaxy velocities over what
was possible in Paper I.

Figure 7 shows the central 100 $h_{70}^{-1}$ kpc radius around each group.  These images have been centered on the BGG rather than
the X-ray center to highlight this galaxy, but for the first four groups these positions are roughly the same.  The
multiple components in RXJ0720+71 and RXJ1256+25 can be clearly seen as well as the smaller secondary in RXJ1205+44.
These components appear to be contained within a common halo, and as noted in Paper I, the secondary in 
RXJ0720+71 has a redshift consistent with the group.  If we treat the multiple components as one object, the first four
groups contain one dominant galaxy at their centers, which is also reflected in the relatively large difference in 
magnitude between this galaxy and the next brightest member.  In these four groups, the X-ray emission is centered on
the BGG (Paper II).  The last three groups instead show associations of several bright galaxies.  As mentioned 
previously, RXJ1648+60 contains a string of bright galaxies, at least eight of which are confirmed members, spread over
a radius of $\sim$200 $h_{70}^{-1}$ kpc.  Both RXJ1334+37 and RXJ1347+07 show compact associations of several galaxies seen in 
projection, and our spectroscopy confirms that at least three and two of these galaxies are group members, respectively.
In these latter three groups the difference in magnitude between the first and second ranked members is smaller,
and in at least RXJ1334+37 and RXJ1648+60 the X-ray emission is not peaked on the BGG (as already noted no XMM-Newton data
are available for RXJ1347+07).  These groups may represent a different evolutionary phase, at least as far as the growth 
of the BGG, from the first four groups.
\clearpage
\begin{center}
\epsscale{0.4}
\plotone{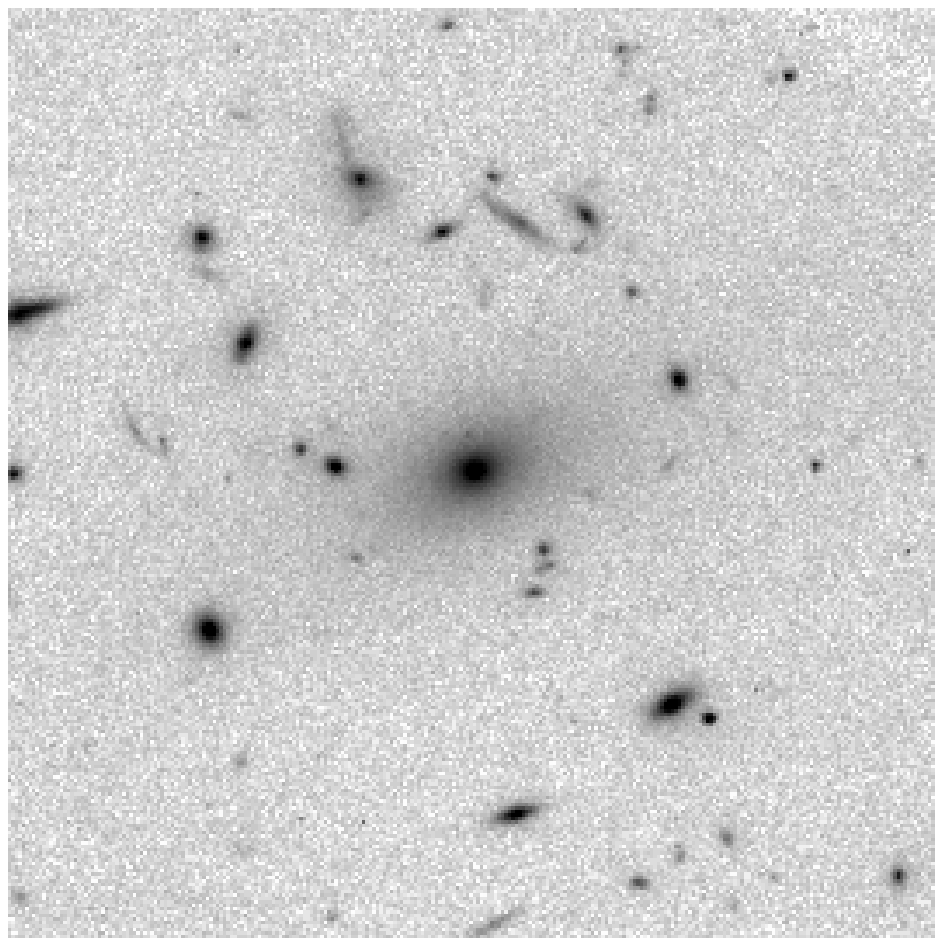}
\plotone{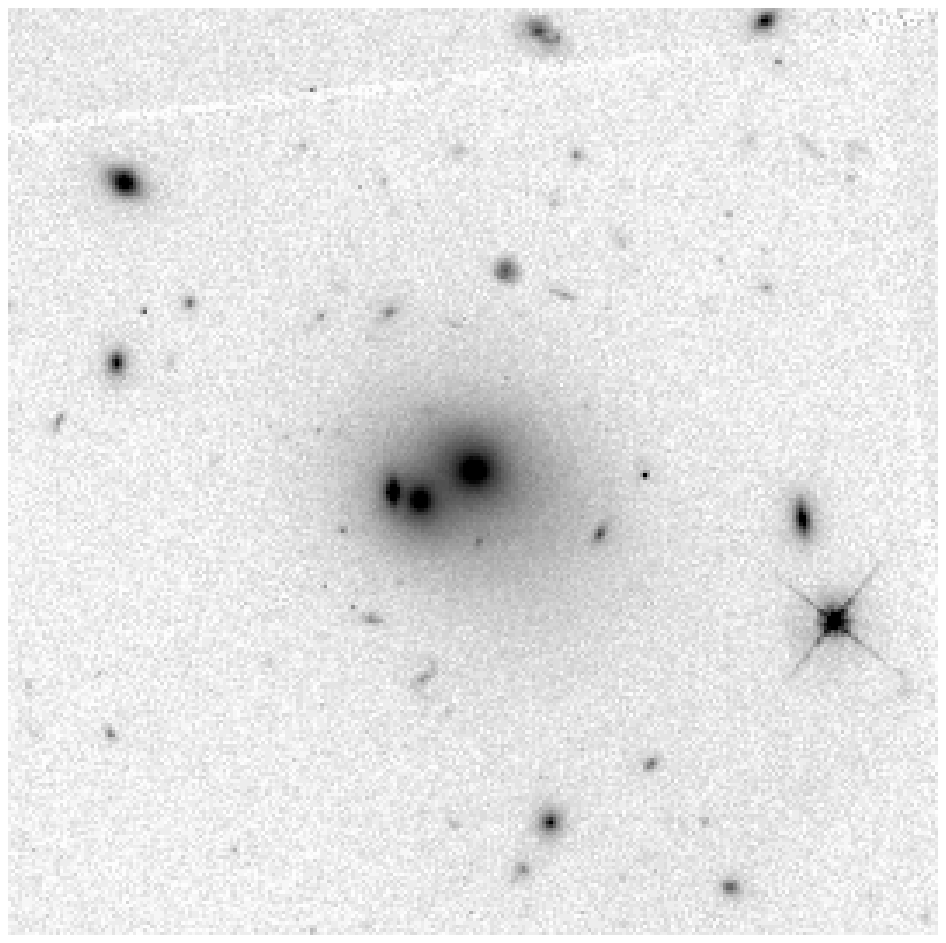}
\plotone{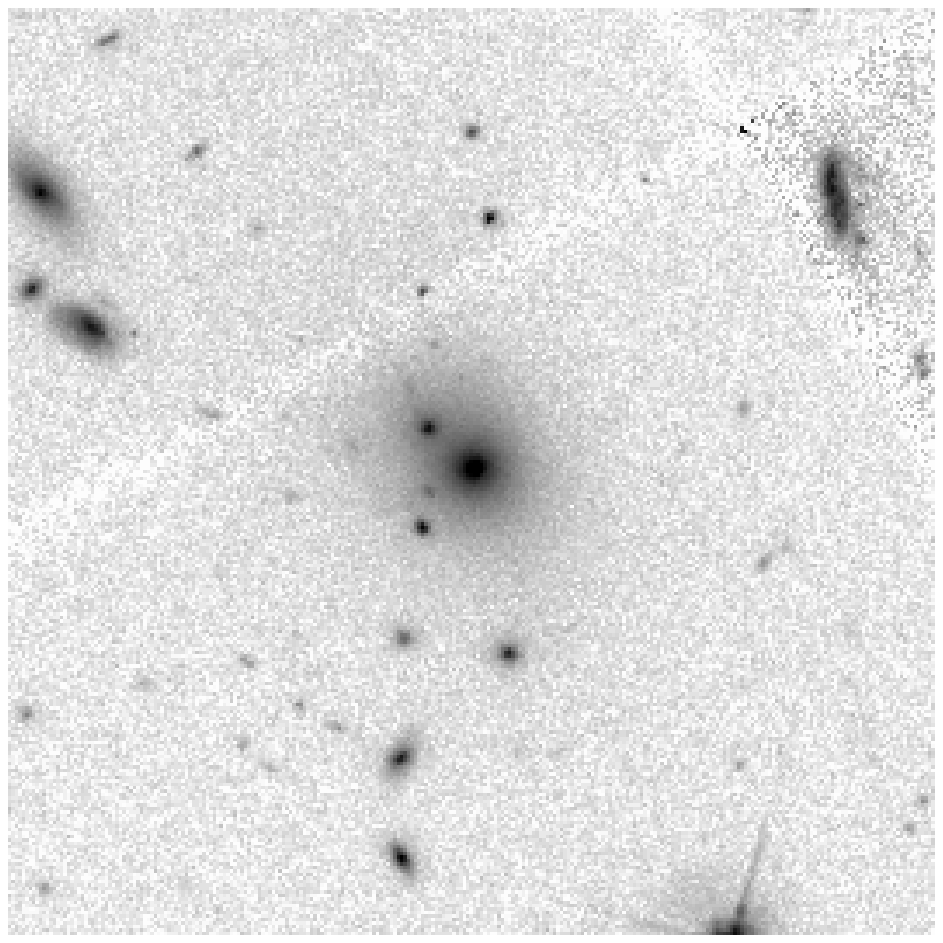}
\plotone{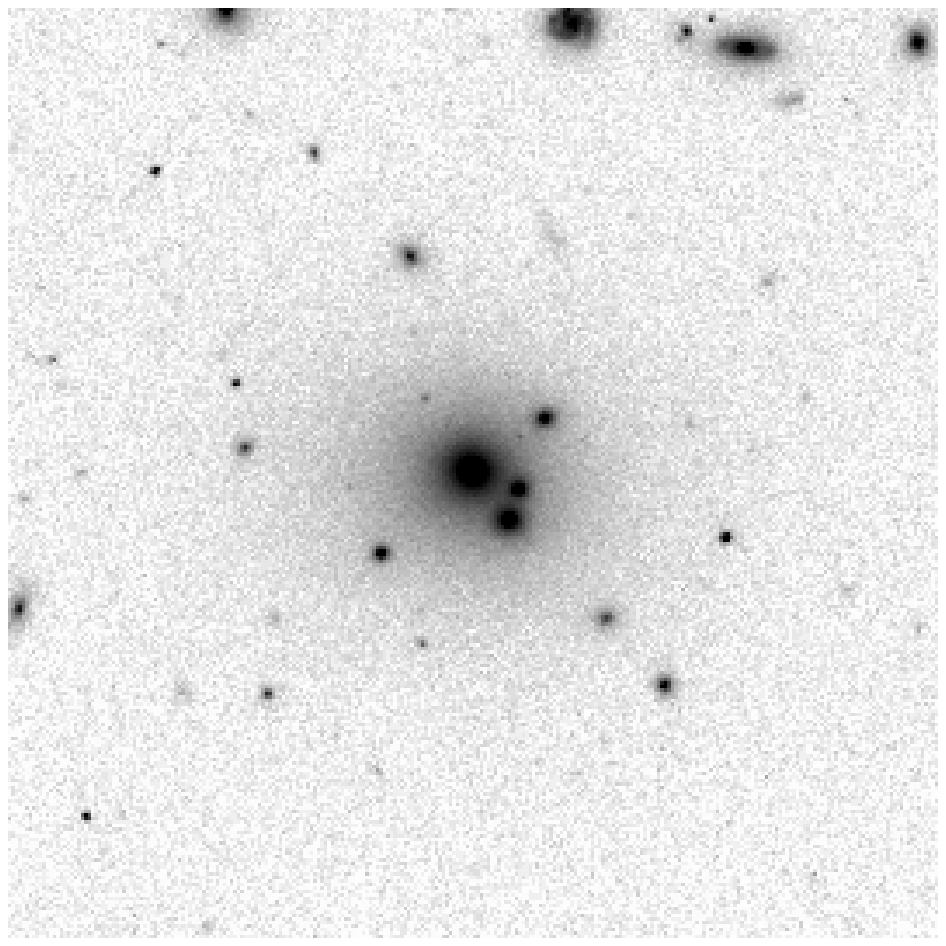}
\plotone{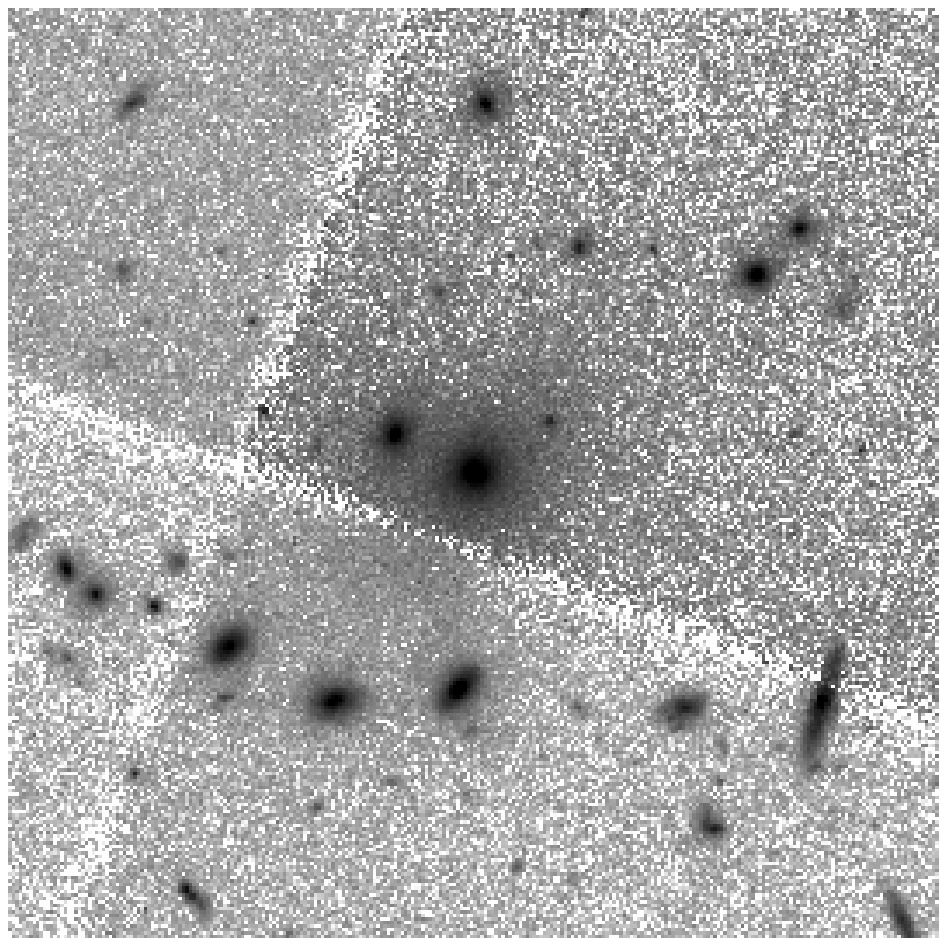}
\plotone{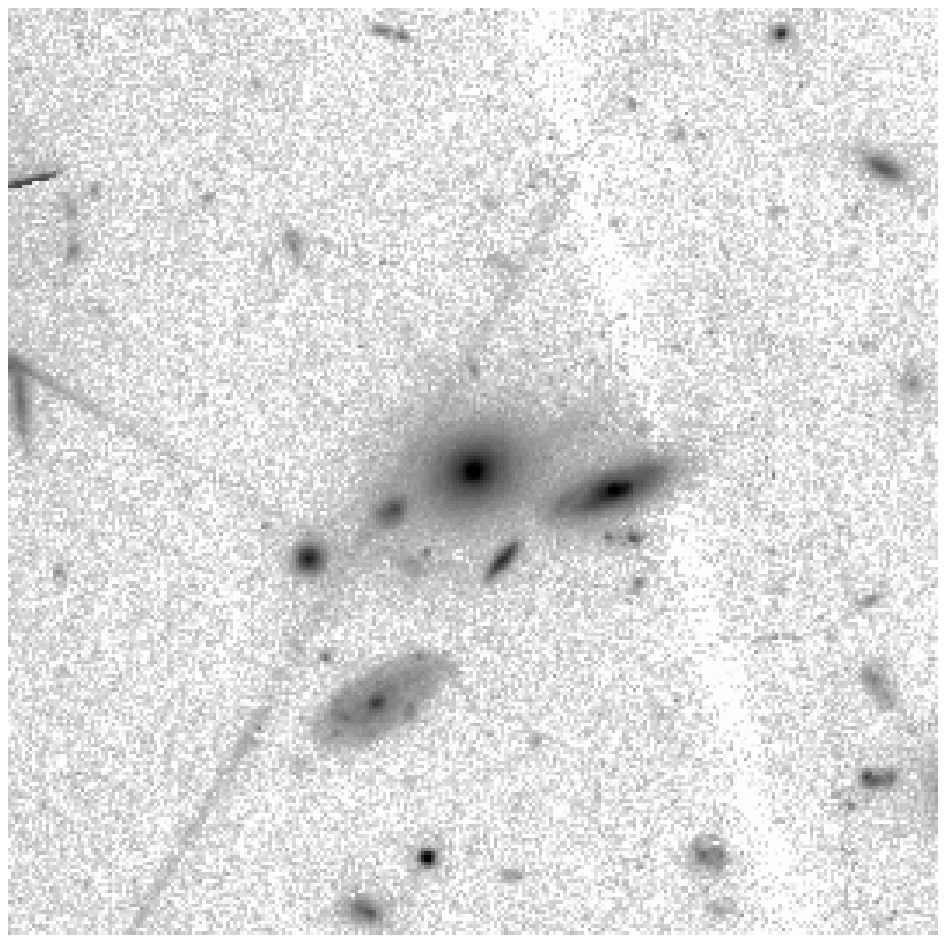}
\end{center}
\clearpage
\begin{figure}
\epsscale{0.4}
\plotone{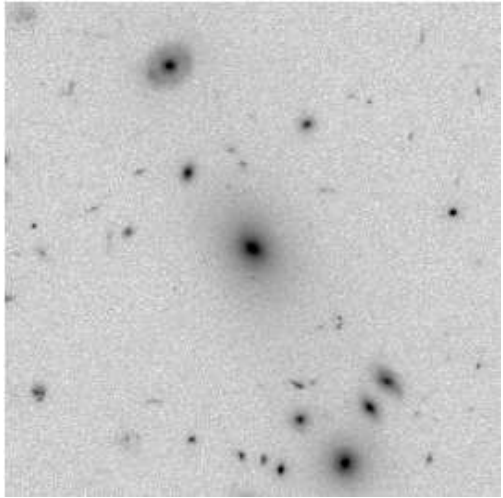}
\caption{ HST images of the central 200x200 $h_{70}^{-1}$ kpc of each group.  These images are centered on the BGG. }
\end{figure}
\clearpage
The observation of relatively compact galaxy associations in these systems leads to a comparison with the compact
groups (CGs) discovered at low-redshift.  CGs, discovered based on imaging, are typically defined as systems
with at least four bright galaxies within 3 mag of the brightest galaxy, with an average surface 
brightness above a fixed level, and closely spaced enough such that the next galaxy as bright is at 3 times the
projected radius or more (e.g. Hickson 1982).  As rich systems, our groups do not meet this final isolation criterion; 
however, a few of them do have several galaxies closely spaced in projection.  The galaxies in CGs typically 
lie within radii of around 100 $h_{70}^{-1}$ kpc (Hickson 1982; Iovino 2002; Iovino et al. 
2003; Pompei, de Carvalho, \& Iovino 2006; Focardi \& Kelm 2002; Barton et al. 1996; Allam \& Tucker 2000; 
Lee et al. 2004).  In column 8 of Table 4, we list the number of galaxies in each group within 100 $h_{70}^{-1}$ kpc of the group 
center and with magnitudes within 3 mag of the BGG.  For RXJ0720+71 and RXJ1256+25, only the BGG itself
meets these criteria, but if the multiple nuclei are considered separately both of these groups have three galaxies in
the core region.  The other groups have between three and eight galaxies in the central 100 $h_{70}^{-1}$ kpc.  We do not have 
complete spectroscopy for these regions, but RXJ0329.0+02 and RXJ1648+60 have four and three confirmed members, respectively,
and we expect on average one field galaxy in a region this size at these magnitudes (Postman et al. 1998).  Therefore, 
at least some X-ray
luminous, intermediate-redshift groups have central densities comparable to CGs.  This result is consistent with the
observation that many CGs at low-redshift are actually associated with larger galaxy systems (Ramella et al. 1994; 
Barton et al. 1996; de Carvalho et al. 1997; Ribeiro et al. 1998; Tovmassian, Plionis, \& Torres-Papaqui 2006).  This 
is also consistent with the fact that many CGs are luminous in X-rays, although typically with lower
luminosities than the groups studied here (Ponman et al. 1996).  
Unfortunately, a similar X-ray selected group sample at low redshift
does not currently exist, so we do not know the frequency of compact galaxy associations in similarly luminous 
low-redshift groups.

We also observe a difference in the number of bright galaxies in our groups.  Column 7 in Table 4 lists the number
of galaxies in each group within a radius of 700 $h_{70}^{-1}$ kpc and brighter than $M_{V}=-22$.  The first four groups have only
one galaxy of this brightness, while RXJ1648+60 has nine.  RXJ1334+37 and RXJ1347+07 also contain more than one very
bright galaxy.  Here we consider only confirmed group members, but even 
including other bright galaxies in the imaging without redshifts the first four groups could not have more than
two to three galaxies of this brightness.  This difference in the number of bright galaxies implies a difference in the 
luminosity functions of these groups, at least at the bright end.  

Within a radius of 200 $h_{70}^{-1}$ kpc, RXJ1648+60 has five 
galaxies brighter than $M_{V}=-22$ and several other bright galaxies that form the central string of galaxies
mentioned previously.  The HST image of this bright string of galaxies is shown in Figure 8, with the eight known
group members circled.  The central Sa galaxy in this string has a large velocity offset more than 3$\sigma$ different
than the mean of the other galaxies.  Excluding this galaxy, the other seven have a velocity dispersion of 
$194^{+66}_{-23}$ km s$^{-1}$.  This velocity dispersion is significantly less than the group as a whole at better 
than 99\%
confidence.  There is not a significant offset between the mean velocity of these galaxies and the group
as a whole.  These galaxies appear to have sunk to the center of the group potential and may eventually merge to 
form the central, dominant elliptical galaxies observed at low redshift and in some of our other groups.  Simple
estimates show that an $L^{*}$ galaxy can fall to the center of a group through dynamical friction within a 
Hubble time (e.g. Zabludoff \& Mulchaey 1998; Merritt 1984).  In addition, galaxy-galaxy mergers are expected to
be most effective in groups, because the velocity of the collision will be similar to the internal galaxy velocity
dispersion (e.g. Zabludoff \& Mulchaey 1998).

For the other
groups we do not have a sufficient number of group members in the central regions to determine accurate velocity
dispersions, but we do note here some of the velocity offsets.  Within the central 100 $h_{70}^{-1}$ kpc, RXJ0329+02 contains the 
BGG and three smaller known group members.  Three of these have velocity offsets from the group mean 
less than 0.5$\sigma$ and all are 
within 1$\sigma$.  In RXJ1334+37, the BGG has an elliptical companion offset in velocity by $\sim60$  km s$^{-1}$;
both of these galaxies lie more than 100 $h_{70}^{-1}$ kpc from the X-ray center.  The other known group member within 100 $h_{70}^{-1}$ kpc in
RXJ1334+37 is an S0 galaxy with a velocity offset more than $1 \sigma$ from this pair.  In RXJ1347+07, there is a group 
spiral that lies close in projection to the BGG but that is significantly offset in velocity.
\clearpage
\begin{figure}
\plotone{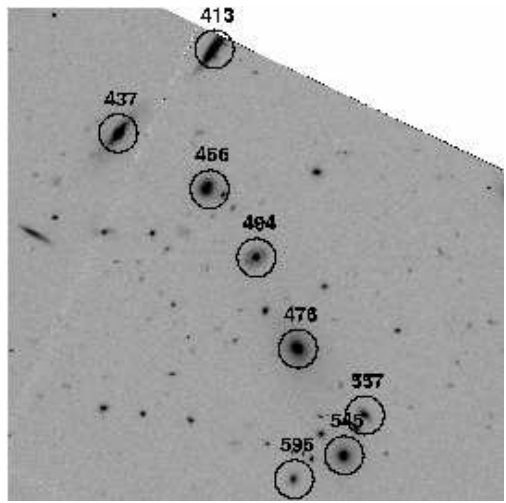}
\caption{ HST image of the central string of bright galaxies in RXJ1648+60. The image is roughly 400x400 $h_{70}^{-1}$ kpc. 
The known group members are circled. }
\end{figure}
\clearpage
Balogh et al. (2002b) find that all 10 of their X-ray selected, low-luminosity clusters ($z\sim0.25$) have giant
elliptical galaxies near the center of the X-ray emission, although their X-ray positions are based on ROSAT data.
These systems are similar in terms of velocity dispersion and redshift to the lowest-redshift groups in our sample. 
Similar to what was found here, they find that none of these central galaxies show [OII] emission.  Examination of 
Figure 5 in Balogh et al. (2002b) and the publicly available HST images reveals that half of these central galaxies
have a close companion or multiple components, although in some cases these secondaries are relatively small.  
In addition, half of the central galaxies lie in compact (in projection at least) associations of several bright 
galaxies.

These observations are consistent with a picture in which BGGs form/grow late with respect to the collapse of the
group itself.  The X-ray observations, high elliptical fractions, and general lack of star-formation all indicate
that these are massive, evolved systems, but a single component, dominant galaxy is not present in most of these
groups.  As noted in Paper I, the groups in our sample appear to represent a range of evolutionary states.  
RXJ1334+37 and RXJ1648+60 appear to be in an earlier stage of evolution.  RXJ1648+60 contains a string of bright
galaxies rather than a single, central galaxy, and based on their velocity dispersion these galaxies lie closer 
to the center of the potential than the rest of the group galaxies.  In RXJ1334+37, the BGG forms a galaxy pair with
another elliptical galaxy, and this pair is offset from the X-ray center.  From the imaging, we also observe many 
other galaxies in the central 100 $h_{70}^{-1}$ kpc of RXJ1334+37.  From the current data, these two groups also have more very bright
galaxies and a smaller magnitude difference between the first and second ranked members than the other groups in
our sample.  In RXJ1347+07, the X-ray center is poorly determined, but the BGG is offset from the center of the
velocity dispersion at fairly high significance (95\% confidence, 10 members).  More data are needed to determine the 
evolutionary state of this group.  It does, however, contain a large number of star-forming galaxies (67\%).  Intriguingly,
these three groups also lie at the low-mass end of our sample in terms of their velocity dispersions, luminosities, and
temperatures.

In the other four groups in our sample the BGG does lie at the center of the X-ray emission, but three of these BGGs 
appear to have multiple components.  In addition, the single-component BGG in RXJ0329.0+02 is fairly elliptical, as 
is the X-ray gas, and there are at least three smaller group members within 100 $h_{70}^{-1}$ kpc of this galaxy.  The two 
BGGs with triple nuclei (RXJ0720+71 and RXJ1256+25) may represent a later stage in the evolution of groups like RXJ1648+60 
and RXJ1334+37 after the central 
galaxies have merged, while RXJ0329+02 and RXJ1205+44 are more relaxed having not undergone a recent major merger.

De Lucia \& Blaizot (2007) use semi-analytical simulations based on halos drawn from the Millennium Simulation 
(Springel et al. 2005) to show that brightest cluster galaxies (BCGs) form relatively late.  They find that
half of the final BCG mass is not accumulated into one galaxy until $z\sim0.5$.  On the other hand, most
of the stars in the BCG progenitors formed early, leading to the old stellar populations in BCGs.  The lack of
significant [OII] emission in our multiple component BGGs supports this picture that central galaxies are built
from old, red galaxies whose mergers do not induce bursts of star-formation.  The central, bright string of
galaxies in RXJ1648+60 contains four early-type galaxies, one E/E merger, one Sa, and one Sb.  The two spiral galaxies
show possible weak [OII] emission (EW[OII]$<$ 4 \AA), but [OII] is not observed in any of the other galaxies.  We
also do not see [OII] in the BGG galaxy pair in RXJ1334+37 or the four galaxies in the central 100 $h_{70}^{-1}$ kpc of 
RXJ0329+02, all of which are early-type galaxies.  This lack of [OII] and the fact that all of these galaxies are
relatively early-types (all earlier than Sc and mostly E and S0 galaxies) indicates that if they did merge to 
form one central galaxy, that these mergers would be
``dry mergers'' that do not induce significant bursts of star formation.  In this scenario, the late formation 
of the BGG is consistent with their observed old stellar ages.

\section{ SUMMARY }

As part of an ongoing program to study the evolution of X-ray luminous groups of galaxies, we obtained 
spectroscopy for seven groups with $0.2<z<0.6$ using the Keck and Gemini telescopes.  From the increased
membership information now available (10-33 members per group), these groups have velocity dispersions
between 200 and 680 km s$^{-1}$, consistent with their X-ray luminosities and temperatures.
This spectroscopy is combined with HST WFPC2 imaging to study morphology and star formation in group galaxies
at $z\sim0.4$.  The properties of the individual groups are summarized in Appendix A.
Our main conclusions are as follows:

1) \textit{X-ray selected groups have galaxy populations similar to clusters at the same redshift.}  In particular, we 
find a large fraction of early-type galaxies ($f_e \sim 70$\%) and a low fraction of galaxies with significant
[OII] emission ($f_{\rm [OII]} \sim 30$\%).  This similarity implies that both the morphological transformation of 
galaxies and the suppression of star formation versus what is observed for field galaxies are effective at 
lower densities than those found in clusters.  Similar to the correlations with radius observed in clusters,
we find that both the fraction of spiral galaxies and the fraction of galaxies with significant [OII] emission
increase with radius from the group center.  However, these fractions remain lower than the field out to at least 
$r_{500}$.  These observations are consistent with observations of clusters that show that differences in 
morphology and star-formation rate persist out past the virial radius and to small local galaxy densities
(Treu et al. 2003; Lewis et al. 2002; Gomez et al. 2003).  We do not, however, observe the difference in velocity
dispersion for early-type versus late-type galaxies or active versus passive galaxies seen in clusters 
by Dressler et al. (1999).

2) \textit{We do not observe a significant evolution in morphology or star formation from the galaxy populations 
observed in low-redshift, X-ray luminous groups.}  A small evolution could be hidden in the errors, but in general
intermediate-redshift groups already have large numbers of early-type galaxies and relatively few star-forming
galaxies.  This lack of evolution apparently contrasts with the evolution in the fraction of star-forming galaxies
observed in optically-selected groups (Wilman et al. 2005b) and clusters (Poggianti et al. 2006), but direct comparison
is hampered by differences in completeness, magnitude limits, and radial sampling.  However, it is clear from 
low-redshift observations that optically-selected groups form a different population than X-ray luminous groups;
they are typically lower-mass systems with more spiral galaxies and in some cases probably still collapsing 
(Rasmussen et al. 2006). A stronger
evolution in the galaxies in these systems would perhaps not be surprising.  The lower-mass systems in Poggianti et al. 
(2006) show significant scatter in their [OII] fractions, but we find that the average [OII] fractions for both their
low-redshift and intermediate-redshift systems are consistent with what we find here.

3) \textit{We observe a significant variation in galaxy properties from group to group.}  While in general these groups
have high elliptical fractions and low [OII] fractions, these fractions vary significantly, and a couple of groups
have galaxy populations similar to the field.  Poggianti et al. (2006) also find significant variation in the [OII]
fractions of the EDisCS groups.  The field-like galaxy populations observed in a couple of groups are surprising among 
such X-ray luminous systems and could
indicate that the evolution of galaxies does not necessarily correlate with the collapse of the system and the 
establishment of a group scale potential.  We do not observe an obvious correlation of galaxy properties with
group velocity dispersion or luminosity (RXJ1347+07 has a low velocity dispersion, but RXJ1205+44 does not),
which would indicate a correlation with system mass, but here the errors
on individual group $f_e$ and $f_{\rm [OII]}$ are large.  Both a larger sample of groups and deeper spectroscopy are needed
to probe the relationship of the individual galaxy populations to other group properties.  Intriguingly, the two 
groups with apparently young galaxy populations are the two highest-redshift groups in our sample, but again a larger
sample is needed to determine if there is a correlation.  These two groups also have the smallest number of confirmed
members, but we determine redshifts for at least 90\% of the galaxies in both groups, so we are not biased toward
emission-line objects.  In RXJ1205+44, the X-ray emission is fairly symmetric and centered on the BGG.  We do not have
an XMM-Newton observation for RXJ1347+07, but its BGG does appear to be offset in velocity from the center of the group.

4) \textit{Comparing the morphological and spectral properties reveals both a population of gas-poor mergers and a 
population of passive spirals.}  In addition to the three BGGs that are observed to have multiple components, we visually 
identify four galaxies as merging or interacting, roughly 10\% of group galaxies.  None of these galaxies, 
including the BGGs, show significant [OII] emission, and at least some of these could be dry mergers.  Similar gas-poor
mergers were observed in an even higher fraction of galaxies in the high-redshift cluster MS1054-03 (van Dokkum et al. 
2000; Tran et al. 2005b).  We also find that 41\% of group spirals lack significant [OII] emission.  Passive spirals
are relatively rare in the field, but they have been found in similar fractions in both massive and low-mass
clusters (Poggianti et al 1999; Balogh et al. 2002b).

In Paper I, it was found that X-ray luminous, intermediate-redshift groups do not all have a dominant elliptical galaxy
at the center of their X-ray emission, in contrast to groups of similar luminosity at low redshift.  In addition, in three
of the four groups that do have a central elliptical we observed multiple components in the BGG, indicating that it may have
recently undergone a merger.  Two of these BGGs contain three luminous components of fairly similar magnitudes and are
unlikely to simply be chance superprojections.  In this paper, we revisit the properties of the BGGs
as well as the central regions of these groups.  As was found for the overall galaxy populations, we find significant
variation in both the number of very bright galaxies and the properties of the BGGs.  Here we summarize our main
observations.

5) \textit{All of the BGGs are elliptical galaxies and none of them show significant [OII] emission.} However, the BGGs in 
at least two groups lie $\sim$ 100 $h_{70}^{-1}$ kpc from the center of the X-ray emission, and in a third group the BGG may be 
significantly offset in velocity from the center of the velocity distribution.  The lack of [OII] emission in the BGGs with 
multiple components could indicate that if these are mergers, they are dry mergers.

6) \textit{At least some of these groups have central galaxy densities comparable to compact groups at low-redshift.}
In addition, one group, RXJ1648+60, is composed of a central bright string of galaxies (see Figure 8).  We confirm that
the galaxies in this string have a lower velocity dispersion than the group as a whole, indicating that they have sunk 
to the center of the potential.

7) \textit{Our groups appear to represent a range of evolutionary states, at least as far as the formation of the central 
galaxy.}  RXJ1334+37 and RXJ1648+60, and possibly RXJ1347+07, appear to be in an earlier stage of evolution.  As mentioned 
above, RXJ1648+60 contains
a string of seven bright galaxies that lie near the group center in both radius and velocity rather than one dominant
galaxy.  In RXJ1334+37, the BGG forms a galaxy pair with another elliptical galaxy, and this pair is offset from the X-ray 
center.  These two groups also have more very bright galaxies and a smaller difference in magnitude between the first and 
second ranked group members than is seen in the rest of the sample.  Intriguingly,
these groups along with RXJ1347+07 lie at the low-mass end of our sample in terms of their velocity dispersions, luminosities, and
temperatures, but a larger sample is needed to establish a firm correlation.  The two BGGs with triple nuclei (RXJ0720+71 and 
RXJ1256+25) may represent a later 
stage of evolution after the galaxies in the central regions of groups like RXJ1334+37 and RXJ1648+60 have merged, while
groups with a single dominant galaxy are even more relaxed, having not undergone a recent major merger.

These observations are consistent with the relatively late formation of the brightest group and cluster galaxies.  This 
scenario is supported by the recent semi-analytical simulations of De Lucia \& Blaizot (2007), who find that BCGs do not
contain half of their final mass until  $z\sim0.5$.  In clusters, mergers among galaxies are unlikely, but this mechanism 
is much more effective in groups (Aarseth \& Fall 1980; Barnes 1985; Merritt 1984, 1985; Zabludoff \& Mulchaey 1998).
The observed old stellar populations of BCGs are maintained if they are built out of the mergers of passive, red galaxies,
and in fact we do not observe significant [OII] emission in our multiple-component BGGs or in the central galaxies in
RXJ1334+37 and RXJ1648+60.

\acknowledgments
We sincerely thank Alan Dressler for his advice on various aspects of the data analysis
including the measurement of equivalent widths and for providing detail on the MORPHS
sample.  We also thank Dan Kelson for a number of useful discussions, and the referee for
the helpful comments on our paper.
Based on observations obtained at the Gemini Observatory, which is operated by the
Association of Universities for Research in Astronomy, Inc., under a cooperative agreement
with the NSF on behalf of the Gemini partnership: the National Science Foundation (United
States), the Particle Physics and Astronomy Research Council (United Kingdom), the
National Research Council (Canada), CONICYT (Chile), the Australian Research Council
(Australia), CNPq (Brazil), and CONICET (Argentina).  J.S.M. acknowledges support from NASA
grants NNG 04-GC846 and NNG 04-GG536 and HST grant G0-08131.01-97A.

\appendix
\section{ NOTES ON INDIVIDUAL GROUPS }

\noindent \textbf{RXJ0329.0+0256} (z=0.41)\\
This group has an elliptical X-ray morphology (ellipticity of 0.37).  The X-ray emission is centered 
on the brightest group galaxy, a single component E5, and the position angle of the BGG and X-ray 
emission are aligned.  This group falls toward the middle of our sample in terms of its velocity 
dispersion, luminosity, and temperature, and it has a high early-type fraction ($f_e = 1.0$ from
 the seven member galaxies imaged with HST).

\noindent \textbf{RXJ0720.8+7109} (z=0.23)\\
The lowest-redshift group in our sample; this group has a symmetric X-ray morphology.  The X-ray
emission is centered on a BGG with three luminous, early-type components.  RXJ0720+71 is 
one of the more massive
groups in terms of its velocity dispersion and temperature, and it has a high early-type fraction 
($f_e = 1.0$ from the fourteen member galaxies imaged with HST).

\noindent \textbf{RXJ1205.9+4429} (z=0.59)\\
The highest-redshift, most massive system in our sample in terms of its velocity dispersion, luminosity, 
and temperature.  It is centered on the brightest elliptical
galaxy, which has a smaller elliptical secondary.  It also has a significantly lower elliptical 
fraction than most of the groups in the sample.

\noindent \textbf{RXJ1256.0+2556} (z=0.23)\\
The other $z\sim0.2$ group in our sample and one of the more massive systems based on its velocity 
dispersion, luminosity, and temperature.  It has a symmetric X-ray morphology, and the X-ray emission is 
centered on the BGG.  The BGG in this group is composed of at least three luminous components.
RXJ1256+25 has a high elliptical fraction and low [OII] fraction, similar to most of the groups.

\noindent \textbf{RXJ1334.0+3750} (z=0.38)\\
One of the lower mass groups in our sample based on its velocity dispersion, luminosity, 
and temperature; it appears to have an asymmetric X-ray distribution.  The BGG is 
an E0 offset from the X-ray peak by over 100 $h_{70}^{-1}$ kpc, and it lies close in both projection and velocity
to another elliptical galaxy.  These galaxies appear to lie in a tail of X-ray emission rather than at the X-ray 
peak (see Figure 6 of Paper II).  There are a number of galaxies close in projection to these two, and at least
one is confirmed to be a group member.  However, RXJ1334+37 has a high elliptical fraction and low [OII] 
fraction, similar to most of the groups.

\noindent \textbf{RXJ1347.9+0752} (z=0.46)\\
We do not have an XMM-Newton observation of this group, but its velocity dispersion places it at the low mass end of our
sample.  The BGG in this group is an E1 galaxy that is offset from the mean velocity of the system at about the
95\% confidence level (based on 10 members).  RXJ1347+07 has a significantly higher [OII] fraction than the rest
of the groups in our sample and a relatively low elliptical fraction.

\noindent \textbf{RXJ1648.7+6019} (z=0.38)\\
This group contains a string of bright galaxies spread over the central 200 $h_{70}^{-1}$ kpc rather than one 
dominant galaxy.  These galaxies have a significantly lower velocity dispersion than the rest of the system of
$\sim 200$ km s$^{-1}$.  Unfortunately, the XMM-Newton observation of this group was highly contaminated with flares, but the
X-ray emission extends over the region covered by the central string of galaxies and does not appear to be peaked
on any one galaxy.  Based on the available XMM-Newton data, this group has a luminosity and temperature at the low-mass end 
of the sample, but its global velocity dispersion is roughly in the middle of the sample.  
Most of these central galaxies are early-types and none of them show significant [OII] emission.
RXJ1648+60 has a high elliptical fraction and low [OII] fraction, similar to most of the groups.

\clearpage
\begin{deluxetable}{lccccccccl}
\tabletypesize{\small}
\rotate
\tablecolumns{10}
\tablewidth{0pt}
\tablecaption{ Spectroscopic Catalog }
\tablehead{
\colhead{RA} & \colhead{Dec} & \colhead{z}    & \colhead{z error} & \colhead{Tel.} & \colhead{R mag.} & \colhead{Mag. error} & \colhead{HST} & \colhead{E/L} & \colhead{Comments}\\
\colhead{(J2000)} & \colhead{(J2000)} & \colhead{}    & \colhead{} & \colhead{} & \colhead{} & \colhead{} & \colhead{morph.} & \colhead{type} & \colhead{}}
\startdata
3:28:58.87 &2:57:58.0 &0.40970 &7e-05 &K &20.43 &0.14 &- &- &- \\
3:29:01.30 &2:57:12.6 &0.41162 &0.00011 &K &20.57 &0.15 &SB0 &E &- \\
3:29:01.71 &2:59:17.9 &0.15879 &0.00021 &K &19.39 &0.09 &- &- &- \\
3:29:01.79 &2:57:50.3 &0.41242 &0.00016 &K &20.93 &0.18 &E1/S0 &E &- \\
3:29:02.34 &2:56:14.5 &0.41019 &0.00011 &K &20.68 &0.16 &S0 &E &possible small E companion \\
3:29:02.35 &2:56:27.1 &0.41132 &0.00013 &K &20.32 &0.14 &E4 &E &-\\
3:29:02.42 &2:54:51.8 &0.54790 &8e-05 &K &20.76 &0.17 &- &- &-\\
3:29:02.48 &3:00:09.1 &0.45081 &0.00010 &K &22.27 &0.34 &- &- &-\\
3:29:02.87 &2:56:23.8 &0.41199 &9e-05 &K &18.97 &0.07 &E5 &E &BGG\\
3:29:03.44 &2:57:32.4 &0.03699 &3e-05 &K &20.21 &0.13 &E1 &E &-\\
3:29:03.57 &2:56:17.6 &0.41181 &0.00010 &K &20.48 &0.15 &E3/S0 &E &-\\
3:29:04.68 &2:57:27.5 &0.41108 &7e-05 &K &20.79 &0.17 &- &- &-\\
3:29:04.69 &2:55:42.7 &0.41434 &0.00021 &K &20.30 &0.14 &S0 &E &-\\
3:29:05.44 &2:55:50.7 &0.15468 &0.00015 &K &19.62 &0.10 &Sb &L &-\\
\enddata
\tablecomments{ Full spectroscopic catalog is available electronically. }
\end{deluxetable}

\begin{deluxetable}{lcccccc}
\tablecaption{ Group Properties }
\tablewidth{0pt}
\tablecolumns{5}
\tablehead{
\colhead{Group} & \colhead{} & \colhead{All Members} & \colhead{} & \colhead{} & \colhead{Gemini/Keck} & \colhead{}\\
\colhead{} & \colhead{N} & \colhead{Redshift} & \colhead{$\sigma_{biwt}$ (km s$^{-1}$)} & \colhead{N} & \colhead{Redshift} & \colhead{$\sigma_{biwt}$ (km s$^{-1}$)}
}
\startdata
RXJ0329.0+0256 &12 &0.4113 &$302^{+93}_{-51}$ &12 &0.4113 &$302^{+93}_{-51}$ \\
RXJ0720.8+7109 &29 &0.2308 &$570^{+70}_{-51}$ &16 &0.2309  &$550^{+99}_{-62}$ \\
RXJ1205.9+4429 &10 &0.5933 &$682^{+160}_{-82}$ &8 &0.5937 &$747^{+201}_{-75}$ \\
RXJ1256.0+2556 &33 &0.2317 &$595^{+63}_{-38}$  &26 &0.2320 &$489^{+54}_{-38}$ \\
RXJ1334.0+3750 &17 &0.3837 &$246^{+44}_{-26}$ &13 &0.3836 &$277^{+53}_{-28}$ \\
RXJ1347.9+0752 &10 &0.4643 &$211^{+57}_{-30}$ &9 &0.4644 &$225^{+72}_{-25}$ \\
RXJ1648.7+6019 &22 &0.3760 &$417^{+118}_{-86}$ &21 &0.3760 &$433 ^{+114}_{-86}$ \\
\enddata
\tablecomments{ Columns 3 and 4 list the group mean redshifts and velocity dispersions determined
from all known members, while columns 6 and 7 list these quantities determined from only those 
galaxies with either Gemini or Keck redshifts. }
\end{deluxetable}

\begin{deluxetable}{lcccccccc}
\tablecaption{ Morphological and Spectral Properties }
\tablewidth{0pt}
\tablecolumns{9}
\tablehead{
\colhead{Group} & \colhead{N} & \colhead{$f_e$} & \colhead{N} & \colhead{$f_e$} & \colhead{N} & \colhead{$f_{\rm [OII]}$} & \colhead{N} & \colhead{$f_{\rm [OII]}$}\\
\colhead{} & \colhead{(All)} & \colhead{(All)} & \colhead{(Comp.)} & \colhead{(Comp.)} & \colhead{(All)} & \colhead{(All)} & \colhead{(Comp.)} & \colhead{(Comp.)}}
\startdata
RXJ0329.0+0256 &7 &1.00 &7 &1.00 &- &- &- &- \\
RXJ0720.8+7109 &14 &1.00 &3 &1.00 &- &- &- &- \\
RXJ1205.9+4429 &10 &0.30 &8 &0.38 &8 &0.38 &7 &0.43 \\
RXJ1256.0+2556 &33 &0.85 &6 &0.83 &25 &0.20 &5 &0.00 \\
RXJ1334.0+3750 &17 &0.88 &11 &0.82 &13 &0.23 &7 &0.29 \\
RXJ1347.9+0752 &10 &0.50 &7 &0.43 &9 &0.67 &6 &0.67 \\
RXJ1648.7+6019 &21 &0.71 &15 &0.67 &21 &0.19 &15 &0.20 \\
\enddata
\tablecomments{ Group early-type fractions and [OII] fractions determined from both all 
available galaxies (All) and only those galaxies brighter than $M_V=-20.5 + 5$log$_{10} h_{70}$ and with radii less 
than 700 $h_{70}^{-1}$ kpc (Comp.). Also listed are the numbers of galaxies from which these quantities were 
determined (N). }
\end{deluxetable}
\clearpage
\thispagestyle{empty}
\begin{deluxetable}{lcccccccl}
\tabletypesize{\small}
\rotate
\tablecaption{ Brightest Group Galaxies }
\tablewidth{0pt}
\tablecolumns{9}
\tablehead{
\colhead{Group} & \colhead{$R_{BGG}$ ($h_{70}^{-1}$ kpc)} & \colhead{Morph} & \colhead{$M_{V,BGG}$} & \colhead{$M_{V,2}$-$M_{V,1}$} &
\colhead{$v_{off}$ (km s$^{-1}$)} & \colhead{$N_{brt}$} & \colhead{$N_{core}$} & \colhead{Comments}
}
\startdata
RXJ0329.0+0256 &9.2 &E5 &-23.2 &1.3 &138 &1 &6 &\\
RXJ0720.8+7109 &6.2 &E0 &-22.3 &0.9 &-182 &1 &1 (3) &BGG with three components \\
RXJ1205.9+4429 &6.1 &E1 &-22.7 &1.2 &-347 &1 &3 &BGG with two components \\
RXJ1256.0+2556 &5.4 &E0 &-23.0 &1.5 &193 &1 &1 (3) &BGG with three components \\
RXJ1334.0+3750 &110.1 &E0 &-22.9 &0.5 &-14 &3 &8 & \\
RXJ1347.9+0752 &0.0 &E1 &-23.1 &0.7 &-135 &2 &4 & \\
RXJ1648.7+6019 &77.9 &E2 &-22.9 &0.1 &-97 &9 &4 & \\
\enddata
\tablecomments{ Columns 2-6 give the properties of the brightest group galaxy.  Column 7 lists the number of galaxies
in each group brighter than $M_{V}=-22$, and column 8 gives the number of galaxies with $R \leq 100$ $h_{70}^{-1}$ kpc within three
magnitudes of the BGG. In column 8, the parentheses show the numbers for RXJ0720+71 and RXJ1256+25 if the multiple 
components are treated separately. }
\end{deluxetable}


\begin{thebibliography}{99}

\bibitem[]{1091}
Aarseth, S. J., \& Fall, S. M. 1980, ApJ, 236, 43

\bibitem[]{1094}
Allam, S., \& Tucker D. 2000, Astron. Nachr., 321, 101

\bibitem[]{1097}
Auger, M. W.,  Fassnacht, C. D., Abrahamse, A. L., Lubin, L. M., \& Squires, G. K. 2006, AJ, submitted

\bibitem[]{1100}
Balogh, M., et al. 2002a, ApJ, 566, 123

\bibitem[]{1103}
Balogh, M., Bower, R. G., Smail, I., Ziegler, B. L., Davies, R. L., Gaztelu, A., \& Fritz, A. 2002b, MNRAS, 337, 256

\bibitem[]{1106}
Balogh, M. L., Morris, S. L., Yee, H. K. C., Carlberg, R. G., \& Ellingson, E. 1999, ApJ, 527, 54

\bibitem[]{1109}
Balogh, M. L., Schade, D., Morris, S. L., Yee, H. K. C., Carlberg, R. G., \& Ellingson, E. 1998, ApJ, 504, L75

\bibitem[]{1112}
Balogh, M. L., Morris, S. L., Yee, H. K. C., Carlberg, R. G., \& Ellingson, E. 1997, ApJ, 488, L75

\bibitem[]{1115}
Barnes, J. 1985, MNRAS, 215, 517

\bibitem[]{1118}
Barton, E., Geller, M. J., Ramella, M., Marzke, R. O., \& da Costa, L. N. 1996, AJ, 112, 871

\bibitem[]{1121}
Beers, T., Flynn, K., \& Gebhardt, K. 1990, AJ, 100, 32

\bibitem[]{1124}
Brough, S., Forbes, D. A., Kilborn, V., Couch, W. 2006, MNRAS, 370, 1223

\bibitem[]{1127}
Butcher, H., \& Oemler, A., Jr. 1984, ApJ, 285, 426

\bibitem[]{1130}
Cid Fernandes, R., Gu, Q., Melnick, J., Terlevich, E., Terlevich, R., Kunth, D., Rodrigues Lacerda, R., \& Joguet, B. 2004, MNRAS, 355, 273

\bibitem[]{1133}
Coleman, G. D., Wu, C.-C., \& Weedman, D. W. 1980, ApJS, 43, 393


\bibitem[]{1139}
de Carvalho, R. R., Ribeiro, A., Capelato, H., \& Zepf, S. 1997, ApJS, 110, 1

\bibitem[]{1142}
De Lucia, G., \& Blaizot, J. 2007, MNRAS, 375, 2

\bibitem[]{1145}
Dressler, A., Oemler, A., Poggianti, B. M., Smail, I., Shectman, S. A., Couch, W. J., \& Ellis, R. S. 2004, ApJ, 617, 878

\bibitem[]{1148}
Dressler, A., Smail, I., Poggianti, B. M., Butcher, H., Couch, W. J., Ellis, R. S.,  \& Oemler, A. 1999, ApJS, 122, 51

\bibitem[]{1151}
Dressler, A., et al. 1997, ApJ, 490, 577

\bibitem[]{1154}
Dressler, A. 1980, ApJS, 42, 565

\bibitem[]{1157}
Eke, V. R., Baugh, C. M., Cole, S., Frenk, C. S., King, H. M., \& Peacock, J. A. 2005, MNRAS, 362, 1233

\bibitem[]{1160} 
Fassnacht, C.~D., \& Lubin, L.~M. 2002, AJ, 123, 627

\bibitem[]{1163}
Focardi, P., \& Kelm, B. 2002, A\&A, 391, 35

\bibitem[]{1166}
Geller, M. J., \& Huchra, J. P. 1983, ApJS, 52, 61

\bibitem[]{1169}
Gomez, P. L., et al. 2003, ApJ, 584, 210

\bibitem[]{1172}
Gunn, J. E., \& Gott, J. R. 1972, ApJ, 176, 1

\bibitem[]{1175}
Heckman, T., et al. 1995, ApJ, 452, 549

\bibitem[]{1178}
Helsdon, S. F., \& Ponman, T. J. 2003, MNRAS, 339, L29

\bibitem[]{1181}
Hickson, P. 1982, ApJ, 255, 382

\bibitem[]{1184}
Hook, I., Jørgensen, I., Allington-Smith, J. R., Davies, R. L., Metcalfe, N., Murowinski, R. G., Crampton, D. 2004, PASP, 116, 425

\bibitem[]{1187}
Hopkins, A. M. 2004, ApJ, 615, 209

\bibitem[]{1190}
Hopkins, A. M., et al. 2003, ApJ, 599, 971

\bibitem[]{1193}
Horner, D. J. 2001, Ph.D. thesis, University of Maryland

\bibitem[]{1196}
Iovino, A. 2002, AJ, 124, 2471

\bibitem[]{1199}
Iovino, A., de Carvalho, R. R., Gal, R. R., Odewahn, S. C., Lopes, P. A. A., Mahabal, A., \& Djorgovski, S. G. 2003, AJ, 125, 1660

\bibitem[]{1202}
Jansen, R. A., Fabricant, D., Franx, M., \& Caldwell, N. 2000, ApJS, 126, 331

\bibitem[]{1205}
Jeltema, T. E., Mulchaey, J. S., Lubin, L. M., Rosati, P., \& B\"{o}hringer, H. 2006, ApJ, 649, 649 (Paper II)

\bibitem[]{1208}
Jones, L. R., Ponman, T. J., Horton, A., Babul, A., Ebeling, H., \& Burke, D. J.  2003, MNRAS, 343, 627

\bibitem[]{1211}
Kauffmann, G., White, S. D. M., Heckman, T. M., Ménard, B., Brinchmann, J., Charlot, S., Tremonti, C., \& Brinkmann, J. 2004, MNRAS, 353, 713

\bibitem[]{1214}
Kauffmann, G., et al. 2003, MNRAS, 346, 1055

\bibitem[]{1217}
Kenicutt, R. C. 1992, ApJ, 388, 310

\bibitem[]{1220}
Kurtz, M. J., \& Mink, D. J. 1998, PASP, 110, 934

\bibitem[]{1223}
Lee, B. C., et al. 2004, AJ, 127, 1811

\bibitem[]{1226}
Lewis, I., et al. 2002, MNRAS, 334, 673

\bibitem[]{1229}
Lilly, S. J., Le Fevre, O., Hammer, F., \& Crampton, D. 1996, ApJ, 460, L1

\bibitem[]{1232}
Madau, P., Pozzetti, L., \& Dickinson, M. 1998, ApJ, 498, 106

\bibitem[]{1235}
Markevitch, M. 1998, ApJ, 504, 27

\bibitem[]{1238}
McCarthy, J. K., et al. 1998, SPIE, 3355, 81

\bibitem[]{1241}
Merritt, D. 1984, ApJ, 276, 26

\bibitem[]{1244}
Merritt, D. 1985, ApJ, 289, 18

\bibitem[]{1247}
Miles, T. A., Raychaudhury, S., Forbes, D. A., Goudfrooij, P., Ponman, T. J., \& Kozhurina-Platais, V. 2004, MNRAS, 355, 785

\bibitem[]{1250}
Momcheva, I., Williams, K., Keeton, C., \& Zabludoff, A. 2006, ApJ, 641, 169

\bibitem[]{1253}
Mulchaey, J. S., Lubin, L. M., Fassnacht, C., Rosati, P., \& Jeltema, T. E. 2006, ApJ, 646, 133 (Paper I)

\bibitem[]{1256}
Mulchaey, J. S., Davis, D. S., Mushotzky, R. F., \& Burstein, D. 2003, ApJS, 145, 39

\bibitem[]{1259}
Mulchaey, J. S., \& Zabludoff, A. 1998, ApJ, 496, 73

\bibitem[]{1262}
Nakata, F., Bower, R. G., Balogh, M. L., \& Wilman, D. J. 2005, MNRAS, 357, 679

\bibitem[]{1265}
Oemler, A. 1992, in Clusters and Superclusters of Galaxies, ed. A. C. Fabian (Dordrecht: Kluwer), 29

\bibitem[]{1268}
Oke, J. B., Cohen, J. G., Carr, M., Cromer, J., Dingizian, A. \& Harris, F. H. 1995, PASP, 107, 375O

\bibitem[]{1271}
Osmond, J. P. F., \& Ponman, T. J. 2004, MNRAS, 350, 1511

\bibitem[]{1274}
Poggianti, B. M., et al. 2006, ApJ, 642, 188

\bibitem[]{1277}
Poggianti, B. M., Smail, I., Dressler, A., Couch, W. J., Barger, A. J., Butcher, H., Ellis, R. S., \& Oemler, A.
 1999, ApJ, 518, 576

\bibitem[]{1281}
Pompei, E., de Carvalho, R. R., \& Iovino, A. 2006, A\&A, 445, 857

\bibitem[]{1284}
Ponman, T. J., Sanderson, A. J. R., \& Finoguenov, A. 2003, MNRAS, 343, 331

\bibitem[]{1287}
Ponman, T., Bourner, P.D.J., Ebeling, H., \& B\"{o}hringer, H. 1996, MNRAS, 283, 690

\bibitem[]{1290}
Postman, M., et al. 2005, ApJ, 623, 721

\bibitem[]{1293}
Postman, M., Lauer, T. R., Szapudi, I., \& Oegerle, W. 1998, ApJ, 506, 33

\bibitem[]{1296}
Postman, M., Lubin, L. M., \& Oke, J. B. 1998, AJ, 116, 560

\bibitem[]{1299}
Postman, M., Lubin, L. M., \& Oke, J. B. 2001, AJ, 122, 1125

\bibitem[]{1302}
Press, W. H., Teukolsky, S. A., Vetterling, W. T., \& Flannery, B. P. 1992, Numerical Recipes in C (2d ed.; New York: Cambridge University Press)

\bibitem[]{1305}
Ramella, M., Diaferio, A., Geller, M. J., \& Huchra, J. P. 1994, AJ, 107, 1623

\bibitem[]{1308}
Rasmussen, J., Ponman, T.~J., \& Mulchaey, J.~S. 2006, MNRAS, 370, 453

\bibitem[]{1311}
Rasmussen, J., Ponman, T.~J., Mulchaey, J.~S., Miles, T.~A., \& Raychaudhury, S. 2006, MNRAS, 373, 653

\bibitem[]{1314}
Ribeiro, A., de Carvalho, R. R., Capelato, H., \& Zepf, S. E. 1998, ApJ, 497, 72

\bibitem[]{1317}
Rosati, P., Della Ceca, R., Burg, R., Norman, R., \& Giacconi, R. 1998, ApJ, 492, L21

\bibitem[]{1320}
Schiminovich, D., et al. 2005, ApJ, 619, L47

\bibitem[]{1323}
Smail, I., Dressler, A., Couch, W. J., Ellis, R. S., Oemler, A., Butcher, H., Sharples, R. M. 1997, ApJS, 110, 213

\bibitem[]{1326}
Springel, V., et al. 2005, Nature, 435, 629

\bibitem[]{1329}
Taylor, J. E., \& Babul, A. 2005, MNRAS, 364, 515

\bibitem[]{1332}
Temporin, S., \& Fritze-von Alvensleben, U. 2006, A\&A, 447, 843

\bibitem[]{1335}
Tonry, J. L., \& Davis, M. 1979, AJ, 84, 1511

\bibitem[]{1338}
Tovmassian, H., Plionis, M., \& Torres-Papaqui, J. P. 2006, A\&A, 456, 839

\bibitem[]{1341}
Tran, K.-V. H., van Dokkum, P., Illingworth, G. D., Kelson, D. D., Gonzalez, A., \& Franx, M. 2005, ApJ, 619, 134

\bibitem[]{1344}
Tran, K.-V. H., van Dokkum, P., Franx, M., Illingworth, G. D., Kelson, D. D., \& F\"{o}rster Schreiber, N. M. 2005, ApJ, 627, L25

\bibitem[]{1347}
Treu, T., Ellis, R. S., Kneib, J.-P., Dressler, A., Smail, I., Czoske, O., Oemler, A., \& Natarajan, P. 2003, ApJ, 591, 53

\bibitem[]{1350}
Turner, E. L., \& Gott, J. R., III. 1976, ApJS, 32, 409

\bibitem[]{1353}
van Dokkum, P. G., Franx, M., Fabricant, D., Illingworth, G. D., \& Kelson, D. D. 2000, ApJ, 541, 95

\bibitem[]{1356}
Walpole, R. E., \& Myers, R. H. 1993, Probability and Statistics for Engineers and Scientists (5th ed.; New York: Macmillan Publishing Company)

\bibitem[]{1359}
Willis, J. P., et al. 2005, MNRAS, 363, 675

\bibitem[]{1362}
Wilman, D. J., Balogh, M. L., Bower, R.G., Mulchaey, J. S., Oemler, A., Carlberg, R. G., Morris, S. L., \& Whitaker, R. J. 2005a, MNRAS, 358, 71

\bibitem[]{1365}
Wilman, D. J., et al. 2005b, MNRAS, 358, 88

\bibitem[]{1368}
Yan, R., Newman, J. A., Faber, S. M., Konidaris, N., Koo, D., \& Davis, M. 2006, ApJ, 648, 281

\bibitem[]{1371}
Zabludoff, A. I., \& Mulchaey, J. S. 2000, ApJ, 539, 136

\bibitem[]{1374}
Zabludoff, A. I., \& Mulchaey, J. S. 1998, ApJ, 496, 39

\end{thebibliography}
\end{document}